%% file: main.tex
\documentclass[acmsmall,screen,submit]{acmart}

\AtBeginDocument{
  \providecommand\BibTeX{{
    \normalfont B\kern-0.5em{\scshape i\kern-0.25em b}\kern-0.8em\TeX}}}

\setcopyright{acmcopyright}
\copyrightyear{2025}
\acmYear{2025}
\acmDOI{}

\acmJournal{TOSEM}
\acmVolume{1}
\acmNumber{1}
\acmArticle{1}
\acmMonth{10}

\usepackage{color}
\usepackage[edges]{forest}
\usepackage{tabularx}
\usepackage{pifont}
\usepackage{subfigure}
\usepackage{tikz}
\usepackage{pgfplots}
\usepackage{pgf-pie}
\usepackage{longtable}
\usepackage{comment}
\usepackage{tcolorbox}
\usepackage{array}

\definecolor{lightcoral}{rgb}{0.94, 0.5, 0.5}
\definecolor{lightgreen}{rgb}{0.56, 0.93, 0.56}
\definecolor{harvestgold}{rgb}{0.98, 0.85, 0.40}
\definecolor{brightlavender}{rgb}{0.75, 0.58, 0.89}
\definecolor{capri}{rgb}{0.0, 0.75, 1.0}
\definecolor{carminepink}{rgb}{0.92, 0.3, 0.26}
\definecolor{celadon}{rgb}{0.67, 0.88, 0.69}
\definecolor{darkpastelgreen}{rgb}{0.01, 0.75, 0.24}
\definecolor{hidden-draw}{gray}{0.9}

\newcommand{\todo}[1]{}
\renewcommand{\todo}[1]{{\color{red} TODO: {#1}}}

\newcommand{\Checkmark}{\ding{51}}     
\newcommand{\XSolidBrush}{\ding{55}}   

\begin{document}

\title{A Systematic Literature Review of Code Hallucinations in LLMs: Characterization, Mitigation Methods, Challenges, and Future Directions for Reliable AI}

\author{Cuiyun Gao}
\affiliation{%
  \institution{Harbin Institute of Technology}
  \city{Shenzhen}
  \country{China}
}
\email{gaocuiyun@hit.edu.cn}

\author{Guodong Fan}
\authornote{Corresponding author.}
\affiliation{%
  \institution{Shandong Agriculture and Engineering University}
  \city{Zibo}
  \country{China}
}
\email{guodong.fan@126.com}

\author{Chun Yong Chong}
\affiliation{%
  \institution{Monash University Malaysia}
  \city{Bandar Sunway}
  \country{Malaysia}
}
\email{chunyong@ieee.org}

\author{Shizhan Chen}
\email{shizhan@tju.edu.cn}
\affiliation{
 \institution{Tianjin University}
   \city{Tianjin}
   \country{China}
   }
\author{Chao Liu}
\email{liu.chao@cqu.edu.cn}
\affiliation{
 \institution{Chongqing University}
   \city{Chongqing}
   \country{China}
   }
\author{David Lo}
\email{davidlo@smu.edu.sg}
\affiliation{
 \institution{Singapore Management University}
    \city{Singapore}
    \country{Singapore}
 }

\author{Zibin Zheng}
\affiliation{%
 \institution{Sun Yat-sen University}
 \city{Guangzhou}
 \country{China}
 \postcode{510006}
}
\email{zhzibin@mail.sysu.edu.cn}

\author{Qing Liao}
\affiliation{%
  \institution{Harbin Institute of Technology}
  \city{Shenzhen}
  \country{China}
}
\email{liaoqing@hit.edu.cn}

\renewcommand{\shortauthors}{Gao et al.}

\input{sections/abstract}



\maketitle

\input{sections/introduction}
\input{sections/background}

\input{sections/methodology}

\input{sections/RQ1}
\input{sections/RQ2}
\input{sections/RQ3}
\input{sections/RQ4}
\input{sections/chanllenges}
\input{sections/discussion}
\input{sections/conclusion}


\bibliographystyle{ACM-Reference-Format}
\bibliography{references}

\end{document}

%% file: sections/abstract.tex
\begin{abstract}
Model hallucination is one of the most critical challenges faced by Large Language Models (LLMs), especially in high-stakes code intelligence tasks. 
As LLMs become increasingly integrated into software engineering tasks, understanding and mitigating hallucination in code becomes essential. In this survey, we provide a systematic review of hallucination phenomena in code-oriented LLMs from four key perspectives. First, we begin by surveying 60 papers to define hallucination in the context of code and summarize its primary causes, such as data noise, exposure bias, and insufficient semantic grounding, while also tracing recent trends in literature across natural language processing (NLP) and software engineering communities. Second, we review model hallucination surveys in a broader span and summarize representative hallucination mitigation strategies, such as knowledge-enhanced generation, constrained decoding, and post-editing.
Third, we review approaches targeted for code intelligence and highlight code-specific challenges that aggravate hallucination, including syntax sensitivity, strict type systems, and dependence on external libraries. Meanwhile, we analyze how emerging code intelligence tasks, e.g., program analysis, symbolic execution, and unit testing, are utilized to detect and mitigate hallucinations. Fourth, we summarize current evaluation benchmarks, ranging from static metrics to dynamic checks, e.g., compilation and execution correctness, and emphasize the need for hallucination-oriented benchmarks. 
Finally, we highlight several key challenges and future directions. We explore the subtle balance between hallucination and generalization, and distinguish hallucinations from ordinary functional errors. As hallucinations often reveal LLMs' issues, we point out the challenges posed by private codebases, where the lack of public data makes hallucinations harder to detect and correct. This paper lays the foundations and outlines the pathways toward reliable AI by focusing on the understanding, evaluation, and mitigation of hallucinations in code LLMs.

\end{abstract}

%% file: sections/introduction.tex
\section{Introduction}

Large language models (LLMs)  demonstrate impressive capabilities in code-related tasks, such as code summarization~\cite{dhulshette2025hierarchical},  generation~\cite{jiang2024self}, and completion~\cite{izadi2024language}. However, LLMs are also prone to producing outputs that are syntactically correct and plausible, yet factually incorrect, non-executable, or inconsistent with the input intent, referred to as \textit{hallucinations}. 
The concept of hallucinations is originally derived from psychology and widely adopted in artificial intelligence to describe the phenomenon where models produce outputs that lack grounding in input or reality~\cite{macpherson2013hallucination}. While hallucinations are often seen as errors to be eliminated, some studies argue that they are an inherent byproduct of generative inference mechanisms~\cite{sinha2023mathematical, xu2024hallucination}. The hallucinations pose significant challenges to the adoption, safety, and reliability of AI systems across various domains.

\begin{figure}
    \centering
    \includegraphics[width=\linewidth]{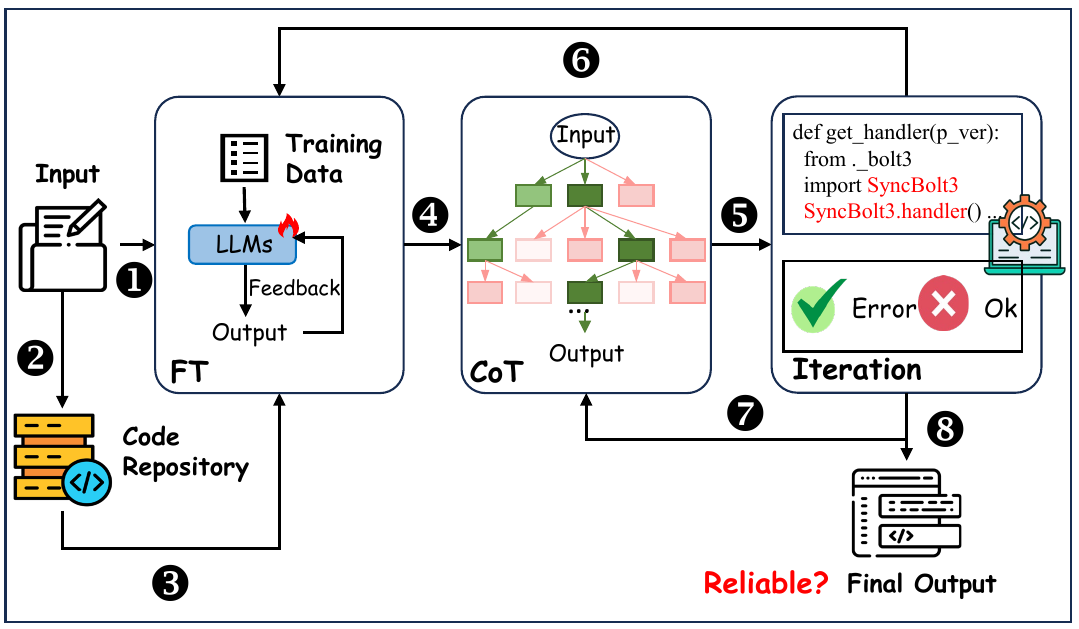}
    \caption{A General Workflow of Code Intelligence. Step 1: User input; Step 2: retrieval-augmented generation incorporates external code and documentation to enrich knowledge; Step 3: model fine-tuning adapts pre-trained models to code-specific tasks and domains; Step 4: chain-of-thought reasoning enhances multi-step reasoning with structured techniques such as tree-of-thought; Step 5: iterative refinement corrects errors and repairs code; Step 6: verification results, corrections, and route back to earlier fine-tuning stages; Step 7: routed back to earlier CoT stages. and Step 8: the system delivers the final output. }
    \label{fig:work_flow}
\end{figure}

Understanding the hallucinations remains a major challenge due to issues such as ambiguity in the field of Natural Language (NL), including programming language (PL). With the rapid development of LLMs, the generated text by LLMs reaches or even surpasses human-level performance on many tasks, e.g., annotation~\cite{wang2024human} and judgment~\cite{wang2025can}. However, hallucinations remain a persistent and serious challenge. For example, models may fabricate citations or invent non-existent objects, significantly undermining user trust in AI systems.
Code-intelligence tasks face the same risk~\cite{zhang2024llm,liu2024exploring}, but with domain-specific stakes. Code is governed by strict syntactic and semantic constraints, depends on modular components, proceeds through multi-stage build/test pipelines, and must execute. These properties reflect a core distinction: natural language targets human interpretation, whereas PLs must be unambiguously parsed and executed by machines while remaining comprehensible to humans~\cite{buse2009learning,allamanis2018survey}. Although PLs have stronger grammatical constraints and are largely characterized by context-free grammars modeled by pushdown automata (PDAs), PLs exhibit stronger domain specificity~\cite{rawte2023survey}. As a result, hallucinations in this context can lead to more severe consequences, such as compilation errors and security vulnerabilities. As shown in Fig.~\ref{fig:work_flow}, from the perspective of the workflow, code intelligence approaches often involve four main stages: retrieval-augmented generation (RAG)~\cite{fan2024survey}, model fine-tuning~\cite{bassamzadeh_comparative_2024}, various chain-of-thought reasoning~\cite{yao2023tree}, and iterative refinement~\cite{eghbali2024hallucinator}.

While many literature reviews exist in the field of NL~\cite{tonmoy2024comprehensive, jiang2024survey, huang2025survey}, there are few literature reviews on hallucination considering code intelligence tasks. 
Lee et al.~\cite{lee2025hallucination} review hallucination by code generation LLMs, categorize types of hallucinations in the code generated by CodeLLMs, review existing benchmarks and mitigation strategies, and identify open challenges. Different from Lee's work, our study offers four advantages. First, we not only examine hallucinations in code generation but also extend our analysis to a wider range of code intelligence tasks, including code summarization, completion, and reasoning. Second, we adopt a cross-disciplinary lens integrating both NLP and software engineering viewpoints. Third, we conduct fine-grained analyses of hallucination causes and manifestations, expanding the discussion to architectural and semantic levels, such as modularity and design pattern violations. Last but not least, our study emphasizes the interplay between mitigation methods and hallucination, offering a more comprehensive understanding of this phenomenon across various tasks.
Overall, we investigate the unique characteristics of PLs and conduct a systematic review of research related to hallucinations. Then, we analyze literature trends, clarify the definition and underlying causes of hallucinations, and summarize representative approaches in both NLP and code intelligence domains. Special attention is given to the unique challenges introduced by the structural and executable nature of code. Finally, we discuss open challenges and future opportunities toward building more reliable code intelligence systems.

To perform a systematic review of this domain, we identified 60 relevant studies from three widely used electronic databases, including ACM Digital Library, IEEEXplore, and Web of Science. 
These databases are selected due to their broad coverage of high-quality publications in the fields of computer science and software engineering.
The search processes involve formulating precise search strings based on key terms related to the hallucinations of LLMs, following the prior studies~\cite{liu2021opportunities, yang2024survey}. We apply a set of inclusion and exclusion criteria to ensure the relevance and quality of the selected studies, with details provided in Section~\ref{sec:method}. The final set of studies serves as the foundation for analyzing the current landscape, identifying trends, and uncovering research gaps in this area.
We aim at investigating the following research questions (RQs) in this study:

\begin{itemize}
    \item \textbf{RQ1: What are the definitions, causes, and literature trends of hallucinations specially in code intelligence?}
    Hallucinations in code refer to outputs that are syntactically correct but semantically incorrect, non-executable, or inconsistent with the input intent. They often stem from insufficient supervision, domain mismatch, or over-generalization. Recent studies attempt to define taxonomy, explore causal mechanisms, and analyze the increasing focus on this phenomenon in code-related tasks.
    \item \textbf{RQ2: How are hallucination mitigation methods in NLP developed and categorized? }
    Various mitigation techniques, such as retrieval-augmented generation (RAG), instruction tuning, and self-consistency decoding, are adapted from NLP to reduce hallucinations. These approaches improve grounding and alignment with intent. However, adapting them to the code domain requires addressing domain-specific
    issues, e.g., structural and execution nature.
    \item \textbf{RQ3: How are hallucination mitigation techniques adapted to address code-specific challenges?}
    Code-specific approaches include syntax-aware decoding, static analysis, execution-based feedback, and test-case validation. Techniques such as constraint-guided generation, compiler-based verification, and structured prompting are used to enforce correctness and reduce unsafe outputs. These methods are often tailored to account for the executable nature of code and strict syntactic semantics.
    \item \textbf{RQ4: What benchmarks and metrics are used to evaluate hallucinations in code intelligence tasks?}
    Several benchmarks, including CodeHaluEval~\cite{tian2025codehalu} and HALLUCODE~\cite{liu_exploring_2024}, provide labeled data for evaluating hallucinations in code generation. Evaluation metrics range from exact match and functional correctness, e.g., pass@k, to hallucination-specific metrics, e.g., API precision and inconsistency rate. Nevertheless, there is a need for unified evaluation protocols and comprehensive multi-domain benchmarks.
\end{itemize}

Based on the findings discussed in all reviewed studies, we observe many challenges in the existing AI systems for code intelligence. These challenges provide opportunities for further research studies, for example:

\begin{itemize}
    \item \textbf{The difference between hallucination and mistake. } Distinguishing hallucinations from ordinary errors remains difficult, as both may manifest as incorrect outputs, but only hallucinations involve confident yet unfounded generation.
    \item \textbf{Confidence vs. Hallucination in LLMs. } High model confidence does not always indicate factual correctness, and overconfidence can increase the likelihood of hallucinations going undetected.
    \item \textbf{The relationship between hallucinations and generalization ability. } While some degree of generalization is necessary for flexible reasoning, it also increases the risk of hallucination when the model extrapolates beyond its training data.
    \item \textbf{The challenge of private data. } Hallucinations are more likely to occur in scenarios involving private codebases or proprietary APIs, where the model lacks sufficient exposure or training data, leading to unsupported or fabricated outputs. 

\end{itemize}

The main contributions of this study are as follows:

\begin{itemize}
    \item We present a cross-domain systematic review of hallucination research spanning NLP and code intelligence, unifying terminology and analyzing representative failure modes.
    \item We summarize representative methods for hallucination detection and mitigation in both NLP and code domains, highlighting their strengths and limitations.
    \item We identify and analyze the unique challenges of hallucinations in code-related tasks, emphasizing the structural and executable nature of PLs.
    \item We highlight challenges and opportunities for research studies, including the distinction between hallucinations and genuine errors, the influence of model confidence, and the tension between generalization and factual accuracy.

\end{itemize}

The remainder of the paper is structured as follows. Section 2 provides background on hallucinations in the context of code generation. Section 3 details the methodology employed in our systematic literature review. Section 4 addresses RQ1 by defining code-related hallucinations, analyzing their primary causes, and reviewing recent literature trends. Section 5 addresses RQ2, presenting mitigation strategies that draw upon techniques developed in the NLP domain. Section 6 focuses on RQ3, discussing approaches tailored to mitigate hallucinations specific to code generation tasks. Section 7 addresses RQ4 by summarizing the key tasks, datasets, models, and evaluation metrics used in this field. Section 8 outlines open challenges and future research opportunities. Section 9 discusses threats to the validity of our review. Finally, Section 10 concludes the paper.

%% file: sections/background.tex
\section{Background}

In this section, we provide essential background on the training process of LLMs, the usage of knowledge, and how these factors relate to hallucination in code intelligence tasks. We then introduce a taxonomy of code intelligence tasks to contextualize the scope of hallucination-related challenges.

\subsection{Training Process and Knowledge Usage for LLMs}
\begin{figure}
    \centering
    \includegraphics[width=\linewidth]{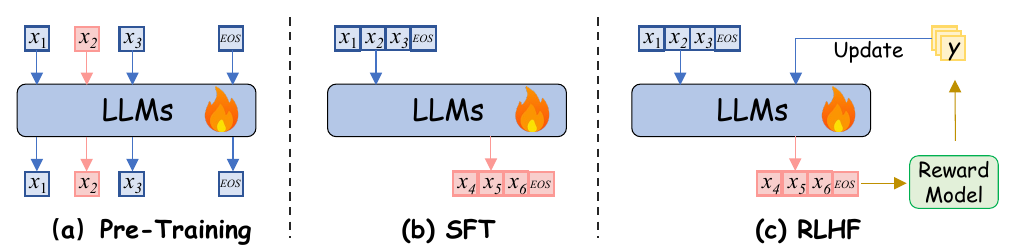}
    \caption{Three key LLM training stages.}
    \label{fig:train_llms}
\end{figure}

The training of LLMs typically involves three key stages, as shown in Fig. \ref{fig:train_llms}, including pre-training, supervised fine-tuning (SFT), and reinforcement learning from human feedback (RLHF). 
In the pretraining stage, the model learns from large-scale, unlabeled text data to understand language patterns and general knowledge~\cite{brown2020language}. Then in the SFT stage, the model is trained on labeled instruction-following data to improve task-specific performance. Finally, RLHF is applied to align the model’s behavior with human preferences, enhancing its helpfulness, safety, and alignment~\cite{ouyang2022training}.
Unlike RLHF, which incorporates negative feedback to suppress undesirable tokens, SFT lacks an explicit signal to indicate which tokens must not be generated. As a result, even with diverse training data, SFT may inadvertently reinforce incorrect tokens, especially when they are frequently annotated or exposed during training.

\begin{figure}[htbp]
    \centering
    \includegraphics[width=\linewidth]{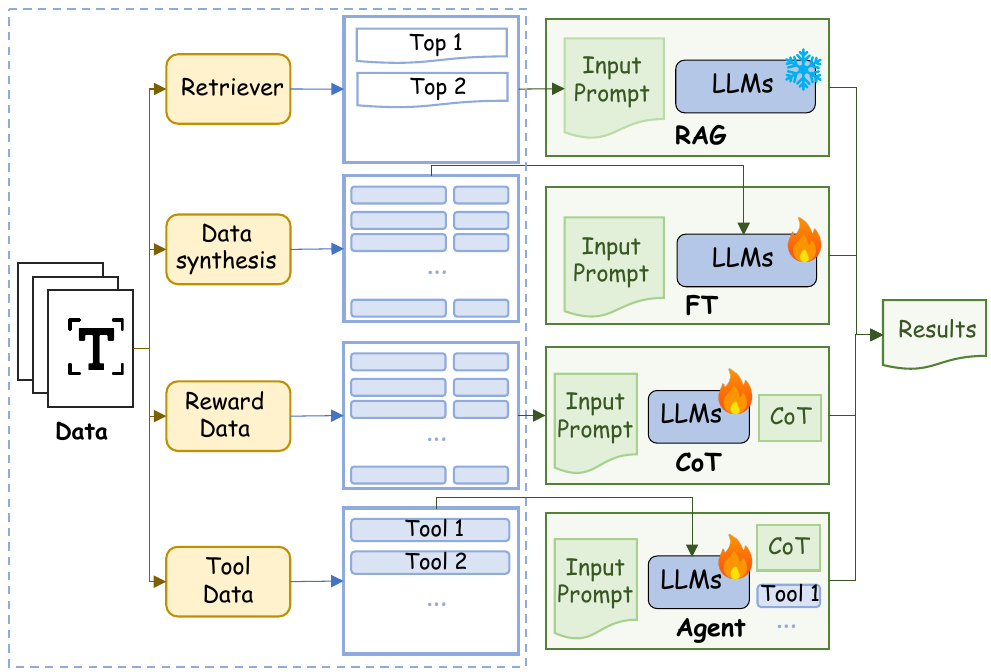}
    \caption{Various Knowledge Usage Techniques.
    }
    \label{fig:external_llm}
\end{figure}

Various techniques are developed to enhance the knowledge utilization capabilities of LLMs, as illustrated in Fig.~\ref{fig:external_llm}.  RAG enables the model to retrieve relevant documents from an external knowledge base and generate responses based on both the input query and the retrieved content~\cite{lewis2020retrieval}. 
Fine-tuning improves task-specific performance by adjusting model parameters using labeled data. While full-parameter fine-tuning is often considered the most effective approach, parameter-efficient alternatives, e.g., P-tuning~~\cite{liu2022p} and LoRA~~\cite{hu2022lora}, are proposed to reduce computational costs while maintaining competitive performance. 
CoT prompting guides the model to produce intermediate reasoning steps, improving performance on complex tasks~\cite{wei2022chain}. In recent years, more advanced reasoning models and approaches have emerged, such as OpenAI’s GPT-4o (o3), DeepSeek-R1, and Tree-of-Thoughts, which further enhance the model's ability to handle multi-step reasoning for decision-making.
Agent-based approaches further extend model capabilities by allowing interaction with external tools or environments to iteratively plan, act, and reason~\cite{yao2023react}.

\subsection{Multi-Perspective of Code Representation}

\begin{figure}
    \centering
    \includegraphics[width=\linewidth]{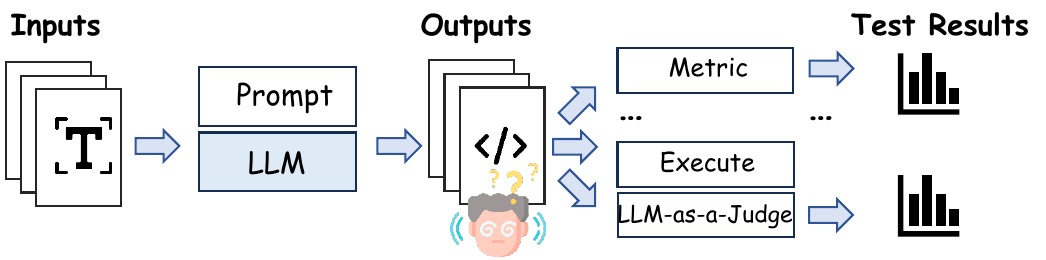}
    \caption{An Illustrative Example of Using LLMs for Code Intelligence Tasks.}
    \label{fig:illustrative}
\end{figure}

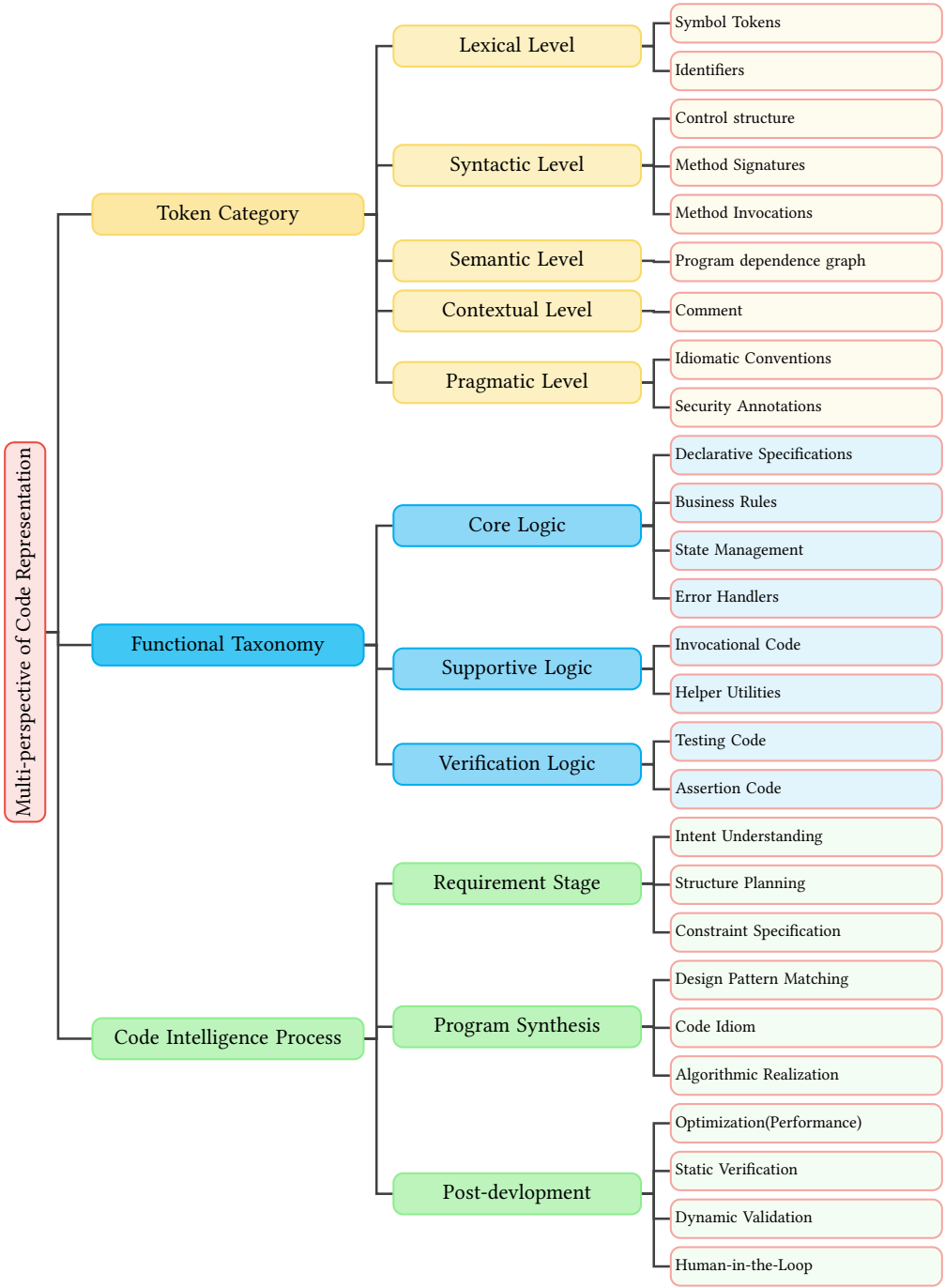
\begin{figure*}[!htb]
    \centering
    \resizebox{\textwidth}{!}{
    \input{figures/aspects}
    }
    \caption{Multi-perspective of Code Representation.}
    \label{fig:aspects_code}
\end{figure*}

Code intelligence aims to empower software development by automating tasks such as code generation~\cite{jiang2024self}, completion~\cite{izadi2024language}, search~\cite{fan2024rapid}, summarization~\cite{dhulshette2025hierarchical}, repair~\cite{yang2024cref}, and translation~\cite{yuan2024transagent}, as shown in Fig. \ref{fig:illustrative}.
To understand the uniqueness of the code-related tasks, it is essential to consider the multi-perspective of core representation: Token Category, Functional Taxonomy, and the Code Intelligence Process, by integrating insights from prior studies~\cite{wang2024natural, kruchten20024+, fu2025ai}, as shown in Fig. \ref{fig:aspects_code}.
\begin{itemize}
    \item The token category defines the granularity at which code can be analyzed and mapped to token vectors~\cite{wang2024natural}. These levels span from the lexical level, e.g., tokens and identifiers, through the syntactic and semantic levels, e.g., grammar structures and Program Dependence Graph (PDG), to the contextual and pragmatic levels, where idiomatic usage, coding conventions, and domain-specific constraints come into play. These layers help to structure code understanding and transformation hierarchically, much like natural language processing.

    \item The functional taxonomy refers to the categorization of code elements based on their intended functions within a software architecture~\cite{kruchten20024+}. It encompasses Core Logic, which forms the main functional body of the code, such as algorithms and data processing routines; Supportive Logic, which includes components like error handling, helper utilities, and invocational code that facilitate the execution of core functions; Verification Logic, which involves testing and assertion code to ensure correctness and reliability. This taxonomy reflects the functional diversity of a codebase, with each category contributing uniquely to the behavior, maintainability, and overall quality of the software.

    \item The code intelligence process outlines the typical lifecycle in DevOps systems~\cite{fu2025ai}. It starts from the requirement understanding stage, where the code captures early-stage planning and user intent, and Program Synthesis, which focuses on the automated generation of code structures. progresses through program synthesis using templates or learned patterns, followed by post-development activities such as optimization, verification, validation, and human-in-the-loop refinement. This process mirrors software engineering workflows to design intelligent code assistants and automation tools.
\end{itemize}

%% file: figures/aspects.tex
\tikzstyle{my-box}=[
    rectangle,
    rounded corners,
    text opacity=1,
    minimum height=1.5em,
    minimum width=5em,
    inner sep=2pt,
    align=center,
    fill opacity=.5,
    color=carminepink!50,
]
\tikzstyle{leaf-1}=[my-box, minimum height=1.5em,
    fill=harvestgold!20, text=black, align=left,font=\scriptsize,
    inner xsep=2pt,
    inner ysep=4pt,
]
\tikzstyle{leaf-2}=[my-box, minimum height=1.5em,
    fill=cyan!20, text=black, align=left,font=\scriptsize,
    inner xsep=2pt,
    inner ysep=4pt,
]
\tikzstyle{leaf-3}=[my-box, minimum height=1.5em,
    fill=lightgreen!20, text=black, align=left,font=\scriptsize,
    inner xsep=2pt,
    inner ysep=4pt,
]
\begin{forest}
    forked edges,
    for tree={
        grow=east,
        reversed=true,
        anchor=base west,
        parent anchor=east,
        child anchor=west,
        base=center,
        font=\large,
        rectangle,
        draw=hidden-draw,
        rounded corners,
        align=left,
        text centered,
        minimum width=4em,
        edge+={darkgray, line width=1pt},
        s sep=3pt,
        inner xsep=2pt,
        inner ysep=3pt,
        line width=0.8pt,
        ver/.style={rotate=90, child anchor=north, parent anchor=south, anchor=center},
    },
where level=0{font=\small}{},
where level=1{text width=11em,font=\small}{},
where level=2{text width=10em,font=\small}{},
where level=3{text width=11em,font=\small}{},
where level=4{text width=11em,font=\small}{},
[Multi-perspective of Code Representation, ver, color=carminepink!100, fill=carminepink!15,
                text=black
  [Token Category , color=harvestgold!100, fill=harvestgold!60, text=black
    [Lexical Level, color=harvestgold!100, fill=harvestgold!40, text=black
    [Symbol Tokens, leaf-1]
    [Identifiers, leaf-1]
    ]
    [Syntactic Level, color=harvestgold!100, fill=harvestgold!40, text=black
    [Control structure, leaf-1]
    [Method Signatures, leaf-1]
    [Method Invocations, leaf-1]
    ]
    [Semantic Level, color=harvestgold!100, fill=harvestgold!40, text=black
    [Program dependence graph, leaf-1]
    ]
    [Contextual Level, color=harvestgold!100, fill=harvestgold!40, text=black
    [Comment, leaf-1]
    ]
    [Pragmatic Level, color=harvestgold!100, fill=harvestgold!40, text=black
      [Idiomatic Conventions, leaf-1]
      [Security Annotations, leaf-1]
    ]
  ]
  [Functional Taxonomy  , color=cyan!100, fill=cyan!60, text=black
    [Core Logic , color=cyan!100, fill=cyan!40, text=black
      [Declarative Specifications, leaf-2]
      [Business Rules, leaf-2]
      [State Management, leaf-2]
      [Error Handlers, leaf-2]
    ]
    [Supportive Logic , color=cyan!100, fill=cyan!40, text=black
      [Invocational Code, leaf-2]
      [Helper Utilities, leaf-2] 
    ]
    [Verification Logic , color=cyan!100, fill=cyan!40, text=black
      [Testing Code, leaf-2]
      [Assertion Code, leaf-2]
    ]
  ]
  [Code Intelligence Process , color=lightgreen!100, fill=lightgreen!60, text=black
      [Requirement Stage , color=lightgreen!100, fill=lightgreen!60, text=black
        [Intent Understanding, leaf-3]
        [Structure Planning, leaf-3]
        [Constraint Specification, leaf-3]  
      ]
      [Program Synthesis , color=lightgreen!100, fill=lightgreen!60, text=black
        [Design Pattern Matching, leaf-3]  
        [Code Idiom, leaf-3]     
        [Algorithmic Realization, leaf-3]  
      ]
      [Post-devlopment , color=lightgreen!100, fill=lightgreen!60, text=black
        [Optimization(Performance) , leaf-3]
        [Static Verification, leaf-3]
        [Dynamic Validation, leaf-3]
        [Human-in-the-Loop, leaf-3] 
      ]
    ]
]
\end{forest}

%% file: sections/methodology.tex
\section{Methodology}\label{sec:method}
In this paper, we conduct a systematic review of the literature according to the guidelines of Kitchenham \cite{kitchenham2004procedures} and Petersen et al. \cite{petersen2015guidelines}, to ensure an unbiased and repeatable procedure. 

\subsection{Research Questions}

To identify, summarize, classify, and analyze the empirical evidence concerning different code intelligence to date, we investigate four research questions (RQs):

\begin{itemize}
    \item \textbf{RQ1:} What are the definitions, causes, and literature trends of hallucinations specially in code intelligence?
    
This question explores how hallucinations are defined in the context of code intelligence, investigates their underlying causes, and analyzes publication trends to understand the evolution of research interest in this area.

    \item \textbf{RQ2:} How are hallucination mitigation methods in NLP developed and categorized? 
    
We examine current strategies for reducing hallucinations, focusing on methods adopted by NLP, e.g., prompt engineering, retrieval augmentation, and reasoning chains, as well as their adaptation to code generation tasks.

    \item \textbf{RQ3:} How are hallucination mitigation techniques adapted to address code-specific challenges?
    
This research question focuses on methods tailored to the unique aspects of code, e.g., API misuse, dependency errors, and environmental inconsistencies, and highlights approaches that go beyond generic NLP mitigation.

    \item \textbf{RQ4:} What benchmarks and metrics are used to evaluate hallucinations in code intelligence tasks? 
    
We survey the existing evaluation protocols, datasets, and quantitative metrics used to measure hallucination frequency and severity, providing insight into the current standards and their limitations.
\end{itemize}

\subsection{Search Strategy}

We identify a set of search terms in code intelligence studies, refine these search terms by checking the titles and abstracts of the relevant papers, combine the terms with the logical ``OR'' and ``AND'', and form the search string for code hallucination studies: \emph{("code generation" OR "code completion" OR "code intelligence" OR "program synthesis") AND ("hallucination" OR "factual inconsistency" OR "faithfulness")}. 
Since there are relatively few studies focusing purely on code-related hallucinations, we turn to the broader field of NLP hallucination research. This extension is justified because both domains share common issues such as factual inconsistency and ensuring the reliability of model outputs. Lessons learned in NLP hallucination research provide methodological insights and mitigation strategies that can inform the study of code hallucinations. 
For these studies, we use the following search string: 
\emph{("hallucination" OR "factual inconsistency" OR "faithfulness") AND ("natural language processing" OR "text generation" OR "language model") AND ("review" OR "survey")
}  
Given the extensive research on hallucinations in NLP, we limit our search to review and survey studies, which provide comprehensive overviews while keeping our focus on code hallucinations. We use the search string to perform an automated search on widely used electronic databases, including ACM Digital Library, IEEExplore, and Web of Science. The search is performed on the title, abstract, and keywords of the papers. We conduct our search on May 25, 2025 and identify the studies published up to that date. 
As shown in Tab. \ref{tab_flow}, we retrieve 71 relevant studies with the automatic search from these three electronic databases. After discarding the duplicated studies, we obtain 60 code search studies. What's more, we manually select representative studies that are not captured by the automatic search, identified through survey reference lists and leading conferences and journals in the field for supplementation. 

\begin{table}[h]
    \centering
    \footnotesize
    \caption{Selection of Related Studies.}
    \begin{tabular}{lcc}
        \toprule
        \textbf{Stage / Source} & \textbf{Code Hallucination} & \textbf{NLP Hallucination} \\
        \midrule
        ACM Digital Library      & 5  & 4  \\
        IEEE Xplore              & 10  & 23 \\
        Web of Science           & 52  & 19 \\
        \midrule
        Total before deduplication      & 67 & 46 \\
        After removing duplicates       & 61 & 46  \\
        After review       & 36 & 24  \\
        \bottomrule
    \end{tabular}
    \label{tab_flow}
\end{table}

\subsection{Study Selection}

Once we retrieve the candidate studies relevant to the code intelligence and survey of LLMs' hallucinations, we assess their relevance based on the following inclusion and exclusion criteria:

\Checkmark \textit{The paper is written in English.}\vspace{3pt}

\Checkmark \textit{The paper involves at least one method that is related to a code intelligence task or offers a survey summarizing existing methods in this area}.\vspace{3pt} 

\XSolidBrush \textit{Keynote records and grey literature are excluded.}\vspace{3pt}

\XSolidBrush \textit{Conference papers that have extended journal versions are excluded.}\vspace{3pt}

\XSolidBrush \textit{Studies that propose new code intelligence methods but do not evaluate their performance are excluded.}\vspace{3pt}

The inclusion and exclusion criteria are piloted by the first and second authors through the assessment of 30 randomly selected primary studies. We measure the reliability of the inclusion/exclusion decisions using Cohen's Kappa statistic \cite{mchugh2012interrater}. The agreement rate in the pilot study is ``moderate'' (0.59), which helps us reach a shared understanding of the criteria. We then apply the criteria to the full set of identified studies. The agreement rate in the full assessment is ``substantial'' (0.73). Disagreements are resolved through open discussions between the first and fourth authors. If a consensus is not reached, the third author acts as a tie-breaker. The entire study selection process takes two weeks. As shown in Tab.~\ref{tab_flow}, we identify 113 candidate studies by reviewing their titles and abstracts. 
We finalize a set of 60 relevant studies through duplication removal, title and abstract review, and full-text screening. 

\subsection{Data Extraction}

To answer the four research questions above, we read the 60 papers and extracted the required data as summarized in Tab. \ref{tab_data}. Our data collection mainly focuses on four kinds of information: publication information, study background, modeling techniques, and benchmark. 

\begin{table}[h]
    \centering
    \footnotesize
    \caption{Extracted Data for Research Questions.}
    \begin{tabular}{llp{7.5cm}@{}}
        \toprule
        \textbf{RQ} & \textbf{Description} & \textbf{Extracted Study Data}\\
        \midrule
        RQ1 & Definition and Publication Trend & Definition, and publication year, publication venue, publication type. \\
        RQ2 & Modeling Techniques &  Empirical study, case study, and Model (type, major technique), auxiliary technique.\\ 
        RQ3 & Modeling Techniques for Code Intelligence & Model tailored for Code. \\
        RQ4 & Benchmark & Tasks, models, dataset, and evaluation metrics. \\
        \bottomrule
    \end{tabular}
    \label{tab_data}
\end{table}

%% file: sections/RQ1.tex
\section{RQ1: definition, causes, and literature trends of hallucinations}


To address RQ1, we first collect relevant studies from multiple sources, which are screened according to the inclusion criteria described in Section 3. We then extract information regarding definitions, causes, and trends of hallucinations, and organize the data to provide a comprehensive overview.

\subsection{Definition of Code Hallucination}

\tcbset{
  colback=gray!5!white,
  colframe=gray!50!black!60,
  boxrule=0.5pt,
  arc=2pt,
  left=6pt,
  right=6pt,
  top=4pt,
  bottom=4pt,
  fonttitle=\bfseries,
}

\begin{tcolorbox}[title=Definition: Code Hallucination, label=def:code_hallucination]
Code hallucination refers to the phenomenon in which 
 LLM-produced outcomes in code intelligence tasks appear  
syntactically correct or even semantically plausible, yet rely on fabricated, non-existent, or context-unsupported elements
~\cite{tian2025codehalu}. 
Unlike a generation error, hallucination arises from the model’s overgeneralization 
and overconfident probabilistic inference under uncertainty.
\end{tcolorbox}

By synthesizing insights from the recent NLP survey~\cite{huang2025survey} and the taxonomy of code hallucinations in the code domain~\cite{liu2024exploring, zhang2024llm}, we propose a unified categorization, as presented in Tab.~\ref{tab:categories}. This taxonomy distinguishes hallucinations based on three key dimensions: \textbf{factuality}, which concerns the correctness of domain and library knowledge; \textbf{faithfulness}, which reflects the model's alignment with task-specific requirements and constraints; and \textbf{compatibility}, which addresses errors arising from environment and dependency mismatches. By identifying granular subtypes such as API knowledge conflicts, functional requirement violations, and environment dependency issues, this taxonomy helps expose diverse failure modes in generated code. 
In the categories of code hallucinations, the factuality and compatibility are closely related as both involve issues impacting code correctness. However, factuality refers to errors intrinsic to the code itself, such as misuse of APIs or syntax mistakes, whereas compatibility concerns cases where the code is logically correct but fails due to environmental factors like missing dependencies or incorrect resource paths.

\begin{table}[ht]
\centering
\caption{Refined Categories of Hallucinations in Code.}
\begin{tabular}{p{3.3cm} p{3.7cm} p{6cm}}
\toprule
\textbf{Category} & \textbf{Example} & \textbf{Explanation} \\
\midrule

\multicolumn{3}{c}{\textbf{I. Knowledge Hallucinations (Factuality)}} \\\midrule
API Misuse & \texttt{requests.get("url", data=\{"k": "v"\})} & Hallucinates or misuses real APIs, resulting in incorrect method calls or mismatched parameter types. \\
Library Misuse & \texttt{np.linear\_regression(X, y)} & The model assumes library functionalities that do not exist. \\
Language Feature Misuse & \texttt{async for line in file:} & Reflects a misunderstanding of language semantics or domain constructs. \\

\addlinespace[0.5em]
\midrule
\multicolumn{3}{c}{\textbf{II. Functional Misalignment (Faithfulness)}} \\\midrule
Functional Requirement Violation & Generates bubble sort instead of merge sort & Generated code does not fulfill the functional specification. \\
Non-functional Requirement Violation & Produces an O($n^2$) algorithm for a performance-critical task & Code violates quality attributes such as performance or maintainability. \\
Logic Flow Violation & \texttt{if x > 0: return x else: return -x} & The code’s control logic deviates from intended behavior, causing subtle functional defects. \\

\addlinespace[0.5em]
\midrule
\multicolumn{3}{c}{\textbf{III. Environment and Dependency Hallucinations (Compatibility)}} \\\midrule
Environment Conflicts & Uses \texttt{torch.compile()} in PyTorch < 2.0 & Model fails to match the system or deployment constraints. \\
Dependency Hallucination & \texttt{openai.ChatAPI} assumed to exist & Assumes dependencies that are incompatible or nonexistent. \\
Resource Path Errors & Writes to \texttt{/tmp/output.txt} without checking the directory & Hallucinated code expects external resources that do not exist. \\

\bottomrule
\end{tabular}
\label{tab:categories}
\end{table}

\begin{figure}[h]
    \centering
    \small
    \subfigure[Publication Venue Types.]{
    \begin{tikzpicture}[scale=0.5]
    \pie[rotate=330, text=pin, color={gray!80, gray!40, gray!10}]{30/Conference (11), 16/Journal (6), 54/Arvix(20)}
    \end{tikzpicture}}
    \subfigure[Contribution types.]{
    \begin{tikzpicture}[scale=0.5]
    \pie[rotate=0, text=pin, color={gray!80, gray!40, gray!10, gray!5}]{58/New Methods (21), 27/Empirical Study (10), 14/Benchmark(5), 1/Survey (1)}
    \end{tikzpicture}}
    \caption{Publication Venues and Contribution Types.}
    \label{fig:veune_type}
\end{figure}
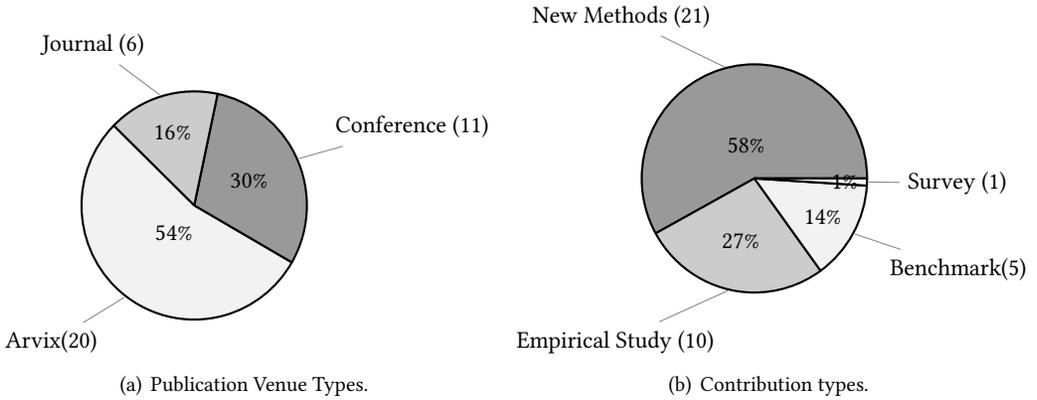

\subsection{The Progress of Hallucination in LLMs}

The concept of hallucination is defined as the perception of an entity or event that is absent in reality, tracing its roots in the fields of pathology and psychology~\cite{macpherson2013hallucination}. For LLMs, hallucination refers to the phenomenon in which generated content is not faithful to the input or appears non-sensical, despite being syntactically fluent~\cite{huang2025survey}.

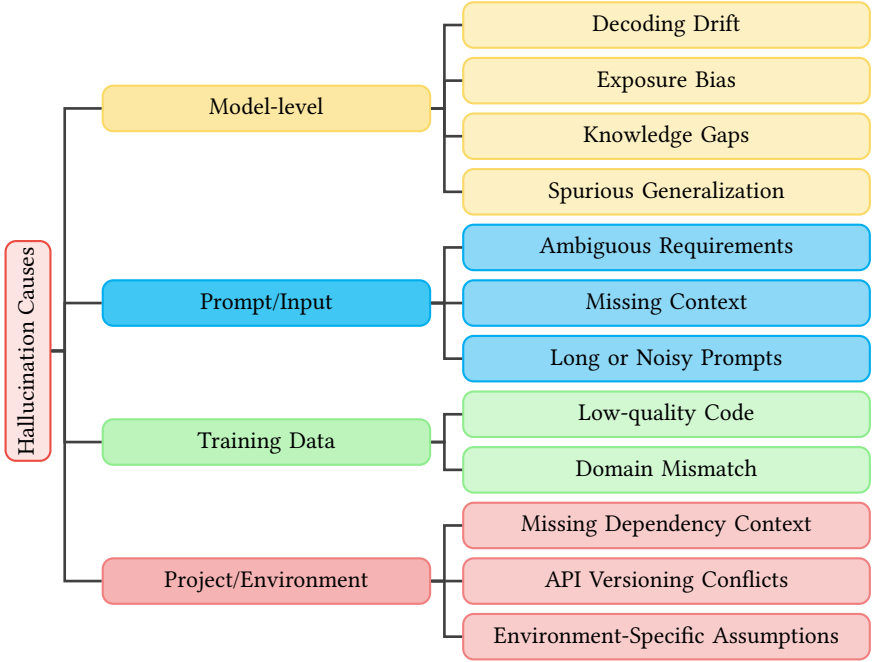
\begin{figure}
    \centering
    \input{figures/taxonomy_hall}
    \caption{The Cause of Hallucinations. }
    \label{fig:taxonomy_hall}
\end{figure}

With the rapid adoption of LLMs, hallucinations have become an issue of rising concern.
LLMs are often incentivized to produce confident outputs rather than express uncertainty, i.e.,  “guessing” behavior, because many evaluation schemes reward answers over abstention or refusal, which helps explain why hallucinated code can appear plausible\cite{kalai2025language}.  
According to previous studies~\cite{huang2025survey, chakraborty2025hallucination, tonmoy2024comprehensive}, hallucination in LLM is attributed to a variety of factors that span model design, ill-formed prompts, data quality, and external execution environment inconsistencies, as shown in Fig.~\ref{fig:taxonomy_hall}. At the model level, issues such as decoding drift, exposure bias, knowledge gaps, and spurious generalization contribute to unfaithful or incoherent outputs. From the perspective of the prompt or input, hallucinations may arise due to ambiguous requirements, missing contextual information, or overly long and noisy prompts. In terms of training data, the use of low-quality or uncurated code corpora and domain mismatch between training and deployment scenarios often exacerbates unreliability. Finally, project- or environment-level factors, including missing dependencies, API version inconsistencies, and implicit environment-specific assumptions, can further lead models to produce code or text that is technically plausible but ultimately unusable. 

\subsection{Trends of LLMs hallucinations}

\begin{figure}[h]
    \centering
    \small
    \subfigure[Number of Publications Per Year for Code Hallucination.]{
    \begin{tikzpicture}
    \begin{axis}[
        ybar, 
        label style={font=\small}, 
        tick label style={font=\small},
        width = 0.4\linewidth,
        height=4cm,
        ymax=23,
        enlargelimits=0.1,
        ylabel={\#Publications},
        y label style={at={(0.05,0.5)}},
        symbolic x coords={2022,2023,2024,2025}, 
        xtick=data, 
        nodes near coords, 
        nodes near coords align={vertical},]
        \addplot[black, fill=gray] coordinates {(2022,0)(2023,5)(2024,20)(2025,13)};
    \end{axis}
    \end{tikzpicture}}
    \subfigure[Number of Publications Per Year for NLP Surveys.]{
    \begin{tikzpicture}
    \begin{axis}[
        ybar, 
        label style={font=\small}, 
        tick label style={font=\small},
        width = 0.4\linewidth,
        height=4cm,
        ymax=20,
        enlargelimits=0.1,
        ylabel={\#Publications},
        y label style={at={(0.05,0.5)}},
        symbolic x coords={2022,2023,2024,2025}, 
        xtick=data, 
        nodes near coords, 
        nodes near coords align={vertical},]
        \addplot[black, fill=gray] coordinates {(2022,0)(2023,6)(2024,12)(2025,6)};
    \end{axis}
    \end{tikzpicture}}
    \caption{Publication trend in years.}
    \label{fig:publication_trend}
\end{figure}
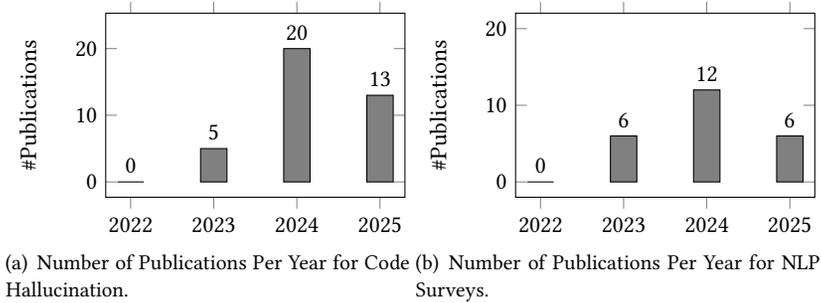

Fig.~\ref{fig:publication_trend}(a) shows the annual number of publications related to code hallucinations from 2023 to 2025. 
We can see a notable rise in 2024 with 20 publications, compared to 5 in 2023. Although 2025 shows a slight decline to 13 publications, it is worth noting that this count only covers the first half of the year. Fig.~\ref{fig:publication_trend}(b) reports the number of NLP-related hallucination survey papers, peaking at 12 in 2024, indicating a growing but still insufficient attention to systematic overviews in this area. 
Fig.~\ref{fig:veune_type}(a) summarizes the venues of code hallucination publications. Preprints on arXiv account for the largest portion (54\%), followed by conferences (30\%) and journals (16\%). 
Although arXiv preprints are sometimes regarded as grey literature, in computer science, they represent a significant portion of impactful and timely research. Therefore, we include them in our review and ensure quality.
This suggests that much of the research is still in early-stage dissemination. In Fig.~\ref{fig:veune_type}(b), we categorize contribution types. A majority (58\%) propose new methods, while empirical studies (27\%) and benchmark creation (14\%) form the rest. Only one survey paper (1\%) exists, further highlighting the lack of comprehensive overviews, particularly in the code intelligence domain.

\begin{center}
    \begin{tcolorbox}
            \textbf{Summary of RQ1:} 
            \begin{enumerate}
           \item Hallucination has gained attention in recent years, overlapping with the period when LLMs become widely adopted.
            \item LLM code hallucinations arise from model flaws, ill-formed prompts, noisy or mismatched training data, and execution environment inconsistencies.
            \item Hallucinations are typically categorized into factual hallucinations (factually incorrect content), faithful hallucinations (content that deviates from the input intent or prompt), and compatibility hallucinations (content is logically correct but fails due to environmental factors).
            \end{enumerate}
    \end{tcolorbox}
\end{center}

%% file: figures/taxonomy_hall.tex
\tikzstyle{my-box}=[
    rectangle,
    rounded corners,
    text opacity=1,
    minimum height=1.5em,
    minimum width=5em,
    inner sep=2pt,
    align=center,
    fill opacity=.5,
    color=carminepink!50,
]
\tikzstyle{leaf-1}=[my-box, minimum height=1.5em,
    fill=harvestgold!20, text=black, align=left,font=\scriptsize,
    inner xsep=2pt,
    inner ysep=4pt,
]
\tikzstyle{leaf-2}=[my-box, minimum height=1.5em,
    fill=cyan!20, text=black, align=left,font=\scriptsize,
    inner xsep=2pt,
    inner ysep=4pt,
]
\tikzstyle{leaf-3}=[my-box, minimum height=1.5em,
    fill=lightgreen!20, text=black, align=left,font=\scriptsize,
    inner xsep=2pt,
    inner ysep=4pt,
]
\begin{forest}
    forked edges,
    for tree={
        grow=east,
        reversed=true,
        anchor=base west,
        parent anchor=east,
        child anchor=west,
        base=center,
        font=\large,
        rectangle,
        draw=hidden-draw,
        rounded corners,
        align=left,
        text centered,
        minimum width=4em,
        edge+={darkgray, line width=1pt},
        s sep=3pt,
        inner xsep=2pt,
        inner ysep=3pt,
        line width=0.8pt,
        ver/.style={rotate=90, child anchor=north, parent anchor=south, anchor=center},
    },
    where level=0{font=\small}{},
    where level=1{text width=12em,font=\small}{},
    where level=2{text width=15em,font=\small}{},
    where level=3{text width=15em,font=\small}{},
    [Hallucination Causes, ver, color=carminepink!100, fill=carminepink!15,
                text=black
      [Model-level,  color=harvestgold!100, fill=harvestgold!60, text=black
        [Decoding Drift,  color=harvestgold!100, fill=harvestgold!40, text=black]
        [Exposure Bias,  color=harvestgold!100, fill=harvestgold!40, text=black]
        [Knowledge Gaps,  color=harvestgold!100, fill=harvestgold!40, text=black]
        [Spurious Generalization,  color=harvestgold!100, fill=harvestgold!40, text=black]
      ]
      [Prompt/Input, color=cyan!100, fill=cyan!60, text=black
        [Ambiguous Requirements, color=cyan!100, fill=cyan!40, text=black]
        [Missing Context, color=cyan!100, fill=cyan!40, text=black]
        [Long or Noisy Prompts, color=cyan!100, fill=cyan!40, text=black]
      ]
      [Training Data , color=lightgreen!100, fill=lightgreen!60, text=black
        [Low-quality Code, color=lightgreen!100, fill=lightgreen!40, text=black]
        [Domain Mismatch, color=lightgreen!100, fill=lightgreen!40, text=black]
      ]
      [Project/Environment, lightcoral!100, fill=lightcoral!60, text=black
        [Missing Dependency Context, lightcoral!100, fill=lightcoral!40, text=black]
        [API Versioning Conflicts, lightcoral!100, fill=lightcoral!40, text=black]
        [Environment-Specific Assumptions, lightcoral!100, fill=lightcoral!40, text=black]
      ]
    ]
\end{forest}

%% file: sections/RQ2.tex
\section{RQ2: Surveys and Mitigation methods in NLP} 

We first analyze the relevant literature reviews. Then, we select the representative detection methods and the mitigation methods. Finally, we examine the benchmark for LLMs' hallucinations in the field of NLP, and highlight the fundamental differences between NL and PL.

\subsection{Surveys for LLMs Hallucinations}
We categorize surveys into the following areas: Foundations and Technical Overviews of LLMs, Applications of Generative AI and LLMs in Healthcare and Clinical Practice, Hallucination: Identification, Evaluation, and Mitigation, and Cross-Domain Applications and Interdisciplinary Integration.

\textbf{Foundations and Technical Overviews of Large Language Models. } Han et al.~~\cite{han2024review} provide a comprehensive review of large language models, covering core technologies, interdisciplinary integration, deployment, and future directions.
Briganti~~\cite{briganti2024chatgpt} reviews ChatGPT’s architecture and applications in healthcare, emphasizing both its transformative potential and current limitations, including a lack of common sense knowledge, a propensity for hallucinating facts, a restricted context window, and potential privacy concerns.
Tonmoy et al.~~\cite{tonmoy2024comprehensive} provide a comprehensive survey of over thirty techniques for mitigating hallucinations in LLMs, introducing a detailed taxonomy based on dataset usage, task types, feedback mechanisms, and retriever types. These analyses highlight the risks of hallucination in sensitive applications and emphasize the importance of grounded generation for real-world deployment.

\textbf{Applications of Generative AI and LLMs in Healthcare and Clinical Practice. } Sallam~~\cite{sallam2023chatgpt} reviews ChatGPT's potential in healthcare education, research, and clinical practice, emphasizing both its benefits and critical concerns such as hallucinations, misinformation, and ethical challenges. 
Chu~~\cite{chu2024chatgpt} reviews ChatGPT’s practical use in veterinary medicine and provides actionable recommendations across clinical, educational, and research contexts.
Cheng~~\cite{cheng2024applications} reviews the applications and challenges of using LLMs in pathology, emphasizing both their potential and the need for cautious implementation.
Wong et al.~~\cite{wong2024review} systematically review the use of LLMs in ophthalmology, noting strong performance in clinical tasks but recurring challenges such as hallucination and domain specificity.
Volkmer et al.~~\cite{volkmer2024large} review the use of LLMs in psychiatry, noting both their clinical promise and challenges, such as bias, privacy, and misinformation risks.
Collectively, these studies demonstrate the growing integration of generative AI in various areas of healthcare, while consistently underscoring domain-specific challenges such as hallucinations, misinformation, and the need for cautious, ethically grounded deployment.

\textbf{Hallucination: Identification, Evaluation, and Mitigation.} Hallucination in NLG has attracted increasing attention due to its negative impact on the reliability of downstream tasks such as summarization, dialogue generation, and question answering. Ji et al.~~\cite{ji2023survey} present a comprehensive survey that systematically reviews definitions, evaluation metrics, mitigation techniques, and task-specific challenges related to hallucinations in NLG, offering insights and future research directions to improve generation faithfulness. 
Huang et al.~~\cite{huang2025survey} provide a comprehensive taxonomy and analysis of hallucinations in LLMs, highlighting unique challenges in open-ended IR systems and discussing both mitigation strategies and future research opportunities.
Despite the widespread use of hallucination in AI, recent studies have pointed out inconsistencies in its definition across domains. For example, Maleki et al.~~\cite{maleki2024ai} conduct a systematic review across fourteen databases and call for a unified terminology to support cross-domain understanding and evaluation.
Agarwal et al.~~\cite{patil2024large} explore hallucination issues in code generation, proposing a taxonomy and benchmark to assess model reliability in high-stakes applications systematically.
Frank et al.~~\cite{frank2024efficient} examine the risks and ethical considerations of using generative AI in academic writing, with a focus on hallucination, bias, and publisher policies.
These works collectively reveal that hallucination is a pervasive yet inconsistently defined challenge across domains, calling for unified taxonomies, standardized benchmarks, and domain-specific mitigation strategies to ensure the reliability and safety of generative AI.

\textbf{Cross-Domain Applications and Interdisciplinary Integration. }
Katwe et al.~~\cite{katwe2023methodical} highlight the growing role of abstractive summarization in biomedical decision support, noting both its promise and challenges, particularly hallucination control and faithfulness in domain-specific applications.
Li et al.~~\cite{li2023survey} systematically review recent advancements in improving factual consistency in summarization, spanning error types, dataset faithfulness, evaluation metrics, and end-to-end modeling.
Abdullahi et al.~~\cite{abdullahi2025time} conduct a systematic review of time-series LLMs, highlighting challenges such as hallucination and long-range dependency modeling while offering a detailed roadmap for future research.
Pester et al.~~\cite{pester2024conversational} demonstrate how LLMs can be effectively embedded in immersive learning systems, highlighting advancements in fine-tuning, hallucination mitigation, and human evaluation.
Yigit et al.~~\cite{yigit2025chatbot} provide a systematic review of chatbot development, highlighting the evolution from ML and DL approaches to LLM-based systems, and addressing challenges such as hallucination and privacy. 
Esposito et al.~~\cite{esposito2024beyond} and Hindi et al.~~\cite{hindi2025enhancing} investigate the effectiveness of RAG-enhanced LLMs in risk analysis and legal applications, respectively, demonstrating low hallucination rates and improved actionability.
Majumdar et al.~~\cite{majumdar_beyond_2024} present a comprehensive review of LLM misuse in cybersecurity contexts, covering threats such as deceptive content generation, language-driven social engineering, and the facilitation of cybercrime, including phishing, password cracking, typosquatting, and malware dissemination.
Lee et al.~~\cite{lee2023utilizing} discuss the benefits and risks of using ChatGPT in clinical research, highlighting concerns around hallucinations, data limitations, and scientific integrity.
Taken together, these studies reflect the growing recognition of hallucination as a critical challenge in domain-specific LLM applications, motivating interdisciplinary efforts toward detection, mitigation, and trustworthiness enhancement.

\subsection{Hallucinations Detection}

Hallucination detection serves as an indirect approach to mitigating hallucinations. Identifying and classifying the types of hallucinations enables more targeted interventions and helps prevent the generation of inaccurate content.

\textbf{Estimate the Hallucination. } As LLMs are increasingly applied to decision-making in autonomous systems, it is crucial to design mechanisms that detect and manage hallucinations, ensuring reliable performance in out-of-distribution scenarios~\cite{chakraborty2025hallucination}.
To better quantify hallucinations in in-context learning, recent work proposes a method to estimate the hallucination rate of conditional generative models based on response likelihood under a Bayesian interpretation of the learning proces~\cite{jesson2024estimating}. 
Zhang et al.~\cite{zhang2024language} identify a phenomenon termed ``hallucination snowballing'', where language models like ChatGPT and GPT-4 not only generate initial incorrect answers but further justify them with additional false claims, many of which they can later recognize as mistakes.

\textbf{Benchmark for Detection. } Many studies have also proposed benchmarks to systematically evaluate hallucination rates. For example, 
Jiang et al.~\cite{jiang2024large} reveal that LLMs may hallucinate even when they possess correct factual knowledge, and propose a method to detect such hallucinations with high accuracy using token probability patterns across model layers.
Jiang et al.~\cite{jiang2024survey} explore how LLM hallucinations, typically seen as limitations, can instead inspire creativity by aligning with divergent thinking and offering novel ideas for future applications. In cognitive science, creativity is often conceptualized from four distinct perspectives: cognitive processes associated with creativity, personal characteristics of creative individuals, creative products or outcomes, and the interaction between the creative individual and the context or environment.
Dziri et al.~\cite{dziri2022origin} investigate whether hallucinations in knowledge-grounded conversational models stem from the training data or model behavior, finding that more than 60\% of benchmark responses contain hallucinations, which in turn causes models to learn and amplify these errors.
Lee et al.~\cite{lee2023mathematical} propose a mathematical framework to define and quantify hallucination and creativity in GPT models, revealing a trade-off between the two and identifying an optimal balance to enhance model performance across tasks.

\subsection{Hallucinations Mitigation}

\begin{figure*}[!htb]
    \centering
    \input{figures/taxonomy_model}
    \caption{Mitigation Methods.}
    \label{fig:taxonomy}
\end{figure*}
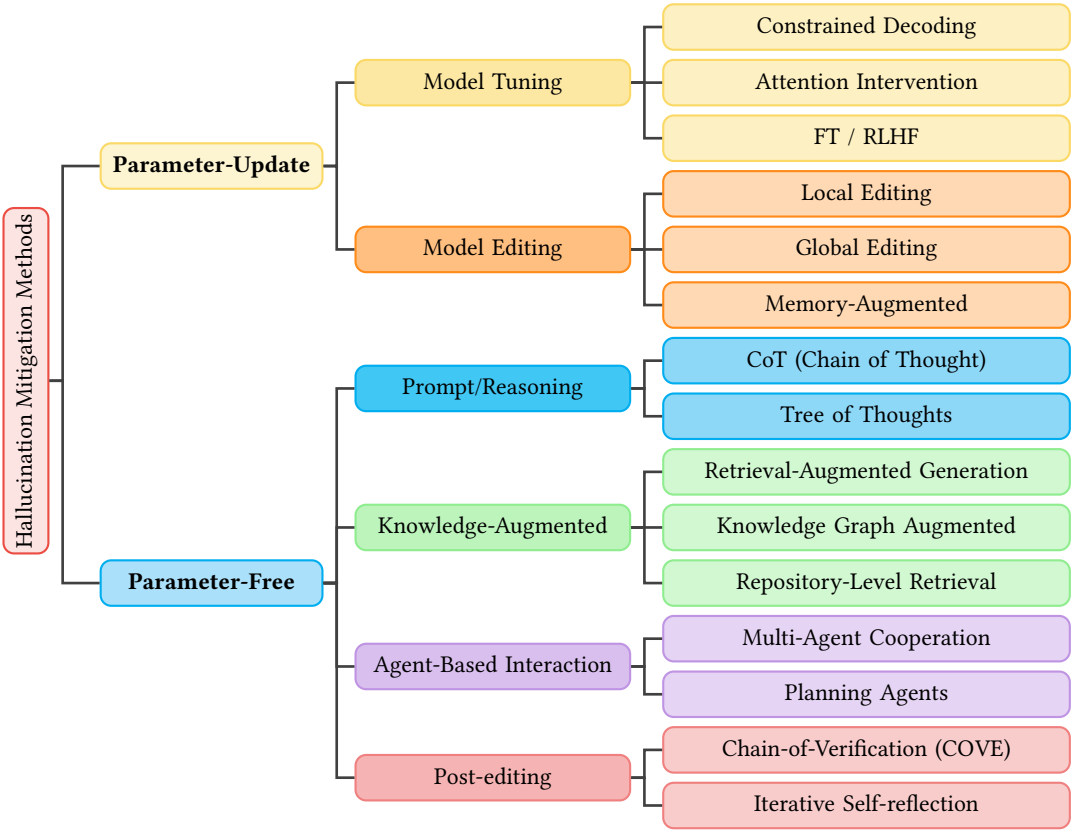

Fig.\ref{fig:taxonomy} presents a categorized overview of existing methods for mitigating code hallucination in LLMs, based on the relevant surveys. These methods fall into five main categories: model-level techniques, e.g., contrastive decoding, attention intervention, prompt and reasoning strategies, e.g., chain of thought, tree of thoughts, knowledge-augmented approaches, e.g., retrieval-augmented generation, agent-based interactions, e.g., multi-agent cooperation, and post-editing decoding, e.g., chain-of-verification, iterative self-reflection. The figure highlights the diversity of strategies but also their alignment with parameter-free versus parameter-update paradigms, offering insights into how different techniques can be flexibly integrated across stages of the generation process.

\subsubsection{Model Tuning for Hallucinations Mitigating}

At the model level, common approaches to mitigating hallucinations include fine-tuning, reinforcement learning alignment, decoding constraints, and attention interventions, with fine-tuning being the most widely used.

\textbf{Constrained Decoding. } Sun et al.~\cite{sun2023contrastive} propose a mixed contrastive objective to explicitly optimize the implicit knowledge elicitation process of LLMs, and thus reduce their hallucination in conversations.
Fu et al.~\cite{fu2024constrained} demonstrate that different decoding methods significantly affect the security of code LLMs, and explore constrained decoding for secure code generation.

\textbf{Attention Intervention.} The attention mechanism is a core element in deep learning that allows the model to selectively focus on relevant parts of an input sequence while downplaying less important information, based on contextual dependencies. Krishna et al.~\cite{krishna2021hurdles} employ sparse attention to improve the model's long-range dependencies for modeling more retrieved documents to mitigate the hallucinations.
Wu et al.~\cite{wu2021controllable} adopt inductive attention, which removes potentially uninformative attention links by injecting pre-established structural information to avoid hallucinations. Nguyen et al.~\cite{nguyen2025hot} introduce Highlighted Chain-of-Thought Prompting (HoT), a method that leverages XML-tagged facts from the input to guide LLMs in grounding their responses. The technique improves attention to instruction-relevant tokens, boosts few-shot performance on 17 tasks, and supports human verification. However, it may also raise concerns about increased user trust in inaccurate results.

\subsubsection{Model Editing for Hallucinations Mitigating}

Model editing emerges as a promising approach to correct factual errors in LLMs without full retraining and can be broadly categorized into three types based on the scope and mechanism~\cite{wang2024knowledge}: \textbf{Local Editing}, \textbf{Global Editing}, and \textbf{Memory-Augmented Editing}. To assess its effectiveness in mitigating real-world hallucinations, Huang et al.~~\cite{huangcan} introduce HalluEditBench, a comprehensive benchmark comprising over 6,000 hallucinations across 9 domains. It evaluates knowledge editing methods across five dimensions: efficacy, generalization, portability, locality, and robustness. Results reveal a significant mismatch between prior benchmark performance and actual hallucination correction, with no method excelling universally. While methods like FT-M and MEMIT perform well on existing datasets, their efficacy drops on HalluEditBench. ICE achieves high scores in some dimensions but lacks robustness, and editing performance is shown to be highly domain- and model-dependent.

\subsubsection{Prompt/Reasoning for Hallucinations Mitigating}

With the emergence of models like Deepseek-Coder R1, an increasing number of studies have recognized the importance of reasoning capabilities in improving generation quality and mitigating hallucinations. 

\textbf{Variants of Chain-of-Thought.} Li et al.~\cite{jiali2025} propose structured CoTs (SCoTs) and present a novel prompting technique for code generation named SCoT prompting, according to which human developers follow structured programming. 
Inspired by the principle of modularization in software development, Pan et al.~\cite{pan2025modularizationbettereffectivecode} propose a prompting technique, called MoT, to enhance the code generation performance of LLMs, which exploits modularization principles to decompose complex programming problems into smaller, independent reasoning steps, enabling a more structured and interpretable problem-solving process.
Zhang et al.~\cite{zhang2024o1} introduce O1-CODER to replicate OpenAI’s o1 model with a focus on coding tasks, which integrates reinforcement learning (RL) and Monte Carlo Tree Search (MCTS) to improve the model's System-2 thinking capabilities.
Arrieta et al.~\cite{arrieta2025o3} assess the safety level of both DeepSeek-R1 (70b version) and OpenAI's o3-mini. After conducting a semi-automated assessment of the outcomes provided by both LLMs, the results indicate that DeepSeek-R1 produces significantly more unsafe responses (12\%) than OpenAI's o3-mini (1.2\%).
Jiang et al.~~\cite{jiang2024self} point out that humans commonly rely on planning to decompose complex problems and sequence solution steps before execution. Inspired by this observation, they incorporate planning into code generation to improve the model's understanding of complex intents and reduce the difficulty of problem-solving.

\subsubsection{Knowledge-enhanced Hallucinations Mitigating}

By incorporating external knowledge through retrieval mechanisms, RAG plays a key role in reducing hallucinations and enhancing content fidelity.
Shuster et al.~\cite{shuster2021retrieval} explore the use of neural-retrieval-in-the-loop architectures for knowledge-grounded dialogue, which can reduce the well-known problem of knowledge hallucination. 
Slobodkin et al.~\cite{slobodkin2024attribute} propose a locally attributable text generation method, named Attribute First, then Generate, which improves factual accuracy and reduces user burden by first selecting relevant source segments and then conditioning generation on them, resulting in more concise, accurate attributions and faster human fact verification in tasks like multi-document summarization and long-form QA.

\subsubsection{Post-editing for Hallucinations Mitigating}

Beyond model training or intervening in the decoding process, post-editing, i.e., refinement of model outputs, can also be employed. Representative approaches include Chain-of-Verification~\cite{dhuliawala2024chain} and self-refinement~\cite{mundlerself}, which iteratively revise generated content to improve correctness and consistency. 
Dhuliawala et al.~\cite{dhuliawala2024chain} develop the Chain-of-Verification (COVE) method whereby the model first drafts an initial response, plans verification questions to fact-check its draft, answers those questions independently so the answers are not biased by other responses, and generates its final verified response. 
Mündler et al.~\cite{mundlerself} point out that an important instance of this problem is self-contradiction, where the language model generates two contradictory sentences within the same context, and present a comprehensive investigation into self-contradiction for various instruction-tuned LMs, covering evaluation, detection, and mitigation. They reveal the prevalence of self-contradictions, e.g., in 17.7\% of all sentences produced by ChatGPT, and then propose a prompting-based framework, through iterations of detect and revise, designed to effectively detect and mitigate self-contradictions. 

\subsubsection{Agent-Based Interaction for Hallucinations Mitigating}

Zhu et al.~~\cite{zhu2025knowagent} propose KnowAgent to mitigate planning hallucinations in LLM-based agents by integrating an explicit action knowledge base and a self-learning strategy to guide action trajectory generation.
Zhou et al.~~\cite{zhou2025guardian} model multi-agent LLM collaboration as a discrete-time temporal attributed graph to detect hallucination propagation and error amplification. It employs an unsupervised encoder-decoder with incremental training to reconstruct graph structures, and uses an information bottleneck-based abstraction mechanism to identify anomalous nodes and edges efficiently and accurately.
Shi et al.~~\cite{shi2025mitigating} introduce a cross-examination framework among agents to verify each reasoning step and filter high-quality responses, addressing hallucinations in complex tasks.

\subsection{Benchmark for LLM Hallucination}

To better understand and mitigate hallucinations in LLMs, researchers have developed a diverse set of benchmarks that evaluate models across multiple dimensions, including factual consistency, functional correctness, security, and domain-specific behavior, covering various domains. 

Li et al.~\cite{li2024dawn} construct the benchmark HaluEval 2.0 to facilitate an empirical study on factual hallucinations in LLMs, focusing on three key research questions: how to detect hallucinations (detection), why LLMs hallucinate (sources), and how to effectively mitigate.
Fu et al.~\cite{fu2024constrained} propose a CODEGUARD+ benchmark, along with two new metrics, to measure the ability of code LLMs to generate both secure and correct code.
Liu et al.~\cite{liu2024exploring} propose a HALLUCODE benchmark to evaluate the performance of code LLMs in the recognition of hallucinations.
Yu et al.~\cite{yu2024codereval} propose a benchmark named CoderEval, consisting of 230 Python and 230 Java code generation problems carefully curated from popular open-source projects in the real world and a self-contained execution platform to automatically assess the functional correctness of the generated code. 
Pal et al.~\cite{pal2023med} propose a benchmark and dataset, Med-HALT (Medical Domain Hallucination Test), designed specifically to evaluate and reduce hallucinations, and show that LLMs are sensitive to prompt framing and decoding parameters, with slight changes often leading to hallucinated responses.
Tian et al.~\cite{tian2025codehalu} introduce the CodeHaluEval benchmark, which includes 8,883 samples from 699 tasks, to systematically and quantitatively evaluate code hallucinations. 
Bang et al.\cite{bang2025hallulens} introduce HalluLens, a unified hallucination evaluation benchmark that combines dynamically generated extrinsic tasks with existing intrinsic tasks to robustly assess LLM hallucinations

\begin{table}[ht]
\centering
\caption{Comparison Between Programming Languages and Natural Languages.}
\label{tab:lang-automata-comparison}
\renewcommand{\arraystretch}{1.3}
\begin{tabular}{@{}p{2.5cm} p{5.5cm} p{5.5cm}@{}}
\toprule
\textbf{Dimension} & \textbf{PL} & \textbf{NL} \\ \midrule
Formality  & High, designed for formal reasoning & Low, naturally evolved \\
Grammar & Mostly context-free languages & Beyond context-sensitive languages \\
Automaton & Pushdown Automaton (PDA) is sufficient & Turing Machine \\
Ambiguity & Nearly unambiguous, easy to parse & Highly ambiguous, requiring contextual understanding \\
Context Dependence & Global and local (e.g., variable scope) & Strong global and pragmatic context required \\
Semantics & Executable semantics with precise meaning & Implicit, often relying on world knowledge \\
Error Tolerance & Low, small errors can cause failure & High, minor errors often tolerated \\
Structure & Strict, hierarchical syntax trees & Flexible, often structurally irregular\\
Parsing Goal & Deterministic syntax and behavior & Interpretation depending on intent and context \\
Model Challenge & Precision, structure enforcement, compilation & Disambiguation, commonsense, reasoning \\ \bottomrule
\end{tabular}
\end{table}

Tab.~\ref{tab:lang-automata-comparison} presents a comparative overview of programming languages (PL) and natural languages (NL) from the perspective of automata theory and language processing. It highlights key differences in formality, grammar complexity, automaton requirements, ambiguity, and context dependence, revealing that while PL emphasizes structural precision and determinism, NL poses greater challenges due to its inherent ambiguity, contextual variability, and semantic flexibility~\cite{linz2022introduction}. PLs are designed for unambiguous interpretation and precise execution, and rely on well-defined grammars, typically context-free, and can be effectively modeled by pushdown automata. In contrast, natural languages (NL) are naturally evolved, inherently ambiguous, and often require more powerful computational models, such as Turing machines, to capture their rich contextual dependencies and semantics.

\begin{center}
    \begin{tcolorbox}
            \textbf{Summary of RQ2:} 
            \begin{enumerate}
            \item Hallucination mitigation is extensively studied in natural language processing, with numerous survey papers summarizing existing methods.
            \item Key mitigation strategies include model-level training, reasoning level, knowledge augmentation, agent-based interaction, and post-editing or decoding constraints.
            \item These strategies exhibit potential applicability to code generation, although adaptations are required to accommodate the structural and semantic properties of code.
        
            \end{enumerate}
    \end{tcolorbox}
\end{center}

%% file: figures/taxonomy_model.tex
\tikzstyle{my-box}=[
    rectangle,
    rounded corners,
    text opacity=1,
    minimum height=1.5em,
    minimum width=5em,
    inner sep=2pt,
    align=center,
    fill opacity=.5,
    color=carminepink!50,
]
\tikzstyle{leaf-1}=[my-box, minimum height=1.5em,
    fill=harvestgold!20, text=black, align=left,font=\scriptsize,
    inner xsep=2pt,
    inner ysep=4pt,
]
\tikzstyle{leaf-2}=[my-box, minimum height=1.5em,
    fill=cyan!20, text=black, align=left,font=\scriptsize,
    inner xsep=2pt,
    inner ysep=4pt,
]
\tikzstyle{leaf-3}=[my-box, minimum height=1.5em,
    fill=lightgreen!20, text=black, align=left,font=\scriptsize,
    inner xsep=2pt,
    inner ysep=4pt,
]

\begin{forest}
    forked edges,
    for tree={
        grow=east,
        reversed=true,
        anchor=base west,
        parent anchor=east,
        child anchor=west,
        base=center,
        font=\large,
        rectangle,
        draw=hidden-draw,
        rounded corners,
        align=left,
        text centered,
        minimum width=4em,
        edge+={darkgray, line width=1pt},
        s sep=3pt,
        inner xsep=2pt,
        inner ysep=3pt,
        line width=0.8pt,
        ver/.style={rotate=90, child anchor=north, parent anchor=south, anchor=center},
    },
    where level=0{font=\small}{},
    where level=1{text width=8em,font=\small}{},
    where level=2{text width=10em,font=\small}{},
    where level=3{text width=15em,font=\small}{},
    [Hallucination Mitigation Methods, ver, color=carminepink!100, fill=carminepink!15, text=black
      [\textbf{Parameter-Update}, color=harvestgold!100, fill=harvestgold!30, text=black
        [Model Tuning, color=harvestgold!100, fill=harvestgold!60, text=black
            [Constrained Decoding, color=harvestgold!100, fill=harvestgold!40, text=black]
            [Attention Intervention, color=harvestgold!100, fill=harvestgold!40, text=black]
            [FT / RLHF , color=harvestgold!100, fill=harvestgold!40, text=black]
        ]
        [Model Editing, color=orange!90, fill=orange!50, text=black
            [Local Editing, color=orange!90, fill=orange!30, text=black]
            [Global Editing, color=orange!90, fill=orange!30, text=black]
            [Memory-Augmented, color=orange!90, fill=orange!30, text=black]
        ]
      ]
      [\textbf{Parameter-Free}, color=cyan!100, fill=cyan!30, text=black
        [Prompt/Reasoning, color=cyan!100, fill=cyan!60, text=black
            [CoT (Chain of Thought), color=cyan!100, fill=cyan!40, text=black]
            [Tree of Thoughts, color=cyan!100, fill=cyan!40, text=black]
        ]
        [Knowledge-Augmented , color=lightgreen!100, fill=lightgreen!60, text=black
            [Retrieval-Augmented Generation, color=lightgreen!100, fill=lightgreen!40, text=black]
            [Knowledge Graph Augmented, color=lightgreen!100, fill=lightgreen!40, text=black]
            [Repository-Level Retrieval, color=lightgreen!100, fill=lightgreen!40, text=black]
        ]
        [Agent-Based Interaction, brightlavender!100, fill=brightlavender!60, text=black
            [Multi-Agent Cooperation, brightlavender!100, fill=brightlavender!40, text=black]
            [Planning Agents, brightlavender!100, fill=brightlavender!40, text=black]
        ]
        [Post-editing, lightcoral!100, fill=lightcoral!60, text=black
            [Chain-of-Verification (COVE), lightcoral!100, fill=lightcoral!40, text=black]
            [Iterative Self-reflection, lightcoral!100, fill=lightcoral!40, text=black]
        ]
      ]
]
\end{forest}

%% file: sections/RQ3.tex
\section{RQ3: Mitigating code-Specific issues} 

In this section, we conclude the code-specific issues for hallucination mitigation. 
While code differs from natural language in its unique syntactic structures and functional characteristics, most existing approaches still adopt approaches similar to those used in natural language processing. 
We present a review of empirical studies on hallucination phenomena, accompanied by a review of detection techniques and mitigation strategies.

\subsection{Empirical Study}

\textbf{Analyzing Code Hallucination. } 
Agarwal et al.~\cite{agarwal2024codemirage} attempt to study hallucinations in the code generated by LLM by introducing the code hallucination definition and a comprehensive taxonomy of code hallucination types, and propose the first benchmark CodeMirage dataset for code hallucinations.
Liu et al.~\cite{liu2024exploring} analyze hallucinations in LLM-generated code, define five main types, and examine their distribution across models and relation to correctness. It introduces HALLUCODE, a benchmark showing that current models struggle to detect and reduce hallucinations.
Zhang et al.~\cite{zhang2024llm} conduct an empirical study to study the phenomena, mechanisms, and mitigation of LLM hallucinations within more practical and complex development contexts in a repository-level generation scenario. 
Lee et al.~\cite{lee2025hallucination} investigate recent studies and techniques relevant to hallucinations generated by CodeLLMs, categorize the types of hallucinations in the code generated by CodeLLMs, review existing benchmarks and mitigation strategies, and identify open challenges.
Huang et al. ~\cite{huang_look_2025} conduct a large-scale exploratory study on risk assessment of LLMs through uncertainty estimation, demonstrating its effectiveness in characterizing nonfactual or uncertain predictions across NLP and code generation tasks.
Currently, there is no unified taxonomy for codes, and hallucination codes are detected through classification models.

\textbf{Code Specific Problems. } The code faces problems such as bugs and choosing third-party libraries. 
LLM-generated code is prone to bugs, Tambon et al.~\cite{tambon2025bugs} examine samples of 333 bugs collected from code generated using three leading LLMs and identify the following 10 distinctive bug patterns: Misinterpretations, Syntax Error, Silly Mistake, Prompt-biased code, Missing Corner Case, Wrong Input Type, Hallucinated Object, Wrong Attribute, Incomplete Generation, and Non-Prompted Consideration. 
Spracklen et al.~\cite{spracklen_we_2024} systematically evaluate package hallucinations in LLM-generated code, pointing out a new type of threat, package hallucinations, to the software supply chain. They reveal the widespread presence of this threat and propose effective mitigation strategies, thereby highlighting a critical and underexplored threat to the software supply chain.
Other studies explore how LLMs can be applied to specialized tasks in areas such as response consistency evaluation and cybersecurity threat analysis. 
Patwardhan et al.~\cite{patwardhan_automated_2024} formally define response consistency for LLMs and propose a framework for its evaluation, introducing self-validation and cross-model validation methods, with experiments on multiple LLMs using a cybersecurity benchmark.
Majumdar et al.~\cite{majumdar_beyond_2024} investigate how malicious actors exploit LLMs for strategic advantage, highlighting threats such as deceptive content generation, social engineering, and various forms of cybercrime, including phishing and malware propagation.
Rahman et al. ~\cite{rahman2024code} propose HallTrigger, a prompt-based technique to systematically trigger and study LLM code hallucinations, revealing their significant threat to software reliability, even in black-box settings. Overall, these studies underscore the urgent need for robust detection and mitigation strategies, especially when deploying LLMs in real-world or safety-critical environments.

\begin{table}[htbp]
  \centering
  \caption{A Summary of Factors Influencing Code in Empirical Studies.}
  \label{Table7-Factors}
  \resizebox{\linewidth}{!}{
    \begin{tabular}{cllp{7cm}}
    \toprule
    \textbf{Year} & \textbf{Author} & \textbf{Factors} & \textbf{Empirical Projects} \\
    \midrule
    2024 & Agarwal et al.~~\cite{agarwal2024codemirage} & Code hallucination taxonomy & CodeMirage benchmark \\
    2024 & Liu et al.~~\cite{liu2024exploring} & Hallucination types and detection & HALLUCODE benchmark \\
    2024 & Zhang et al.~~\cite{zhang2024llm} & Repository-level hallucination analysis & CoderEval benchmark \\
    2024 & Tambon et al.~~\cite{tonmoy2024comprehensive} & Bug patterns & 333 buggy code samples \\
    2024 & Spracklen et al.~~\cite{spracklen_we_2024} & Package hallucinations & 576k samples across 16 LLMs \\
    2024 & Patwardhan et al.~~\cite{patwardhan_automated_2024} & Response consistency & Cybersecurity benchmark \\
    2024 & Majumdar et al.~~\cite{majumdar_beyond_2024} & Malicious LLM usage & Cyber threat case studies \\
    2025 & Lee et al.~~\cite{lee2025hallucination} & Survey and classification & CodeLLMs hallucination types \\
    2025 & Huang et al.~~\cite{huang_look_2025} & Uncertainty estimation & NLP and code risk assessment \\
    \bottomrule
    \end{tabular}
  }
\end{table}

\textbf{For other tasks. } There are also other empirical studies, such as those focusing on zero-shot learning, long-context understanding, and tool usage.
Bang et al.~\cite{bang_multitask_2023} propose a framework for quantitatively evaluating interactive LLMs such as ChatGPT using publicly available datasets. They find that ChatGPT outperforms LLMs with zero-shot learning on most tasks and even outperforms fine-tuned models on some tasks. 
Dong et al.~\cite{dong_bamboo_2023} propose BAMBOO, a multi-task long context benchmark, to comprehensively evaluate the long context ability of LLMs. 
Ferino et al. ~\cite{ferino_junior_2025} conduct a systematic literature review on junior developers’ use of LLM-based tools in software engineering, revealing both the perceived benefits and limitations, such as hallucinations and data leakage.

\subsection{Hallucinations Detection}
In response to the unique characteristics of code, many hallucination detection studies have emerged, targeting issues related to variable names, APIs, and other code-specific elements.

Tian et al.~\cite{tian2025codehalu} define and classify code hallucinations into four types: mapping, naming, resource, and logic hallucinations, propose the CodeHalu detection algorithm, and introduce the CodeHaluEval benchmark. 
Tanzil et al. ~\cite{tanzil_chatgpt_2024} survey software engineering practitioners and propose CID, a tool that detects incorrect ChatGPT responses via metamorphic questioning, achieving an F1-score of 0.74–0.75 in identifying inaccuracies during library selection tasks.
Sun et al. ~\cite{sun_classification-based_2024} propose a human-expert-inspired, classification-based framework for HDL code generation using LLMs, which integrates task decomposition, EDA tool simulation, and search-based variation control to reduce hallucinations and improve Verilog correctness.
Krishna et al. ~\cite{krishna_importing_2025} analyze package hallucination behaviors in LLMs across languages and model settings, revealing security vulnerabilities in the software supply chain and proposing metrics and heuristics to mitigate risks and guide secure AI-assisted development.
Jiang et al. ~\cite{jiang_collu-bench_2024} introduce Collu-Bench, a comprehensive benchmark comprising over 13,000 instances for studying code hallucinations in LLMs across code generation and program repair tasks, providing fine-grained features and experimental insights that highlight the challenges in hallucination prediction and localization.
Zhuo et al.~\cite{zhuo_identifying_2025} present the first comprehensive study of API misuse patterns in LLM-generated code, analyzing both method selection and parameter usage across Python and Java. Through extensive manual annotation of 3,892 method-level and 2,560 parameter-level misuses, they develop a novel taxonomy of four distinct API misuse types specific to LLMs. 
Yang et al.~~\cite{yang2025hallucination} introduce MetaQA, a zero-resource hallucination detection approach based on prompt mutation and metamorphic relations, achieving superior performance over SelfCheckGPT across both open- and closed-source LLMs.

\subsection{Hallucination Mitigation}
Recent efforts to mitigate hallucinations in code intelligence can be categorized into five major areas: model-level modifications, prompting and reasoning strategies, knowledge-augmented generation, agent-based decision frameworks, and post-editing. Code exhibits domain specificity, modular design, specialized structure, and executable feedback~\cite{yang2025code}, making hallucination in code intelligence fundamentally different from that in natural language tasks.

\subsubsection{Knowledge-enhanced LLMs hallucination mitigating}

RAG is a commonly used knowledge-enhanced approach in both NLP and code intelligence tasks.

\begin{figure}
    \centering
    \includegraphics[width=\linewidth]{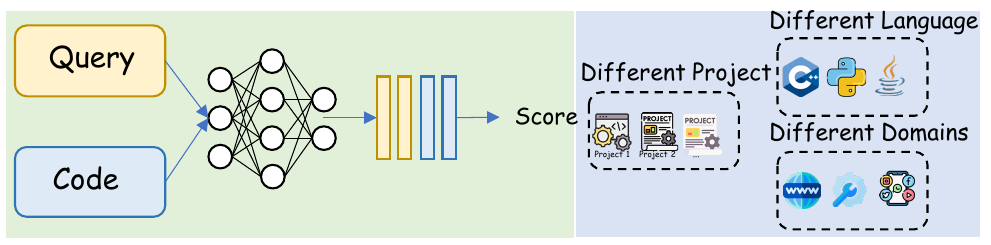}
    \caption{Challenges of code search across domains, projects, and languages: models struggle to generalize beyond the coverage of training data.}
    \label{fig:codesearch}
\end{figure}

\textbf{API Specific Task. }
Eghbali et al.~\cite{eghbali2024hallucinator} present De-Hallucinator, which grounds the predictions of an LLM through a combination of retrieving suitable API references and iteratively querying the model with increasingly suitable context information in the prompt.
Bassamzadeh et al. ~\cite{bassamzadeh_comparative_2024} investigate the challenges of DSL code generation with LLMs due to custom and frequently updated function names, and show through an ablation study that optimized RAG strategies can match fine-tuned Codex models in code similarity while offering better generalization to unseen APIs, despite trade-offs in hallucination and syntax error rates.

The evolution of PLs reflects a shift from low-level control toward higher-level abstraction~\cite{antoy2010functional}. Object-oriented languages incorporate visibility levels and encapsulation for hiding the representation of data and the management of memory. Declarative languages, such as functional and logic languages, enable developers to specify what a program should accomplish rather than how, promoting code clarity, reusability, and efficient execution through formal yet executable specifications.
As shown in Fig.\ref{fig:codesearch}, compared to natural language tasks, RAG in code intelligence involves several domain-specific challenges:
\begin{itemize}
    \item \textbf{Domain specificity.} Code is often tied to specific technical domains, e.g., web, data science, requiring retrieval systems to understand domain-relevant APIs and libraries~\cite{gu2025effectiveness}.

    \item \textbf{Project dependence.} Code usage is highly context-dependent~\cite{bi2024iterative}. Project-specific structure, naming conventions, and dependency configurations must be considered for accurate retrieval.

    \item \textbf{Multi-language complexity.} Real-world codebases frequently span multiple programming languages, e.g., Python, JavaScript, SQL, making retrieval and alignment across heterogeneous syntaxes and semantics more difficult~\cite{zhu2025across}.
\end{itemize}

\textbf{Other Code Scenario. }
Lee et al. ~\cite{lee_novel_2025} propose a method for automatically generating semiconductor equipment control interface code using generative AI, aiming to enhance code completeness and reduce the impact of hallucinations by leveraging existing code bases.
Roychowdhury et al. ~\cite{roychowdhury_eratta_2024} propose ERATTA, an LLM-based extreme RAG system for enterprise-scale sustainability data, featuring multi-LLM orchestration, hallucination detection via five scoring metrics, and structured response generation with >90\% confidence across diverse domains.
Khan et al. ~\cite{khan_rfsensinggpt_2025} present RFSensingGPT, a hybrid RAG framework leveraging filtered RedPajama data and hierarchical chunking for high-fidelity retrieval, achieving a 13\% average faithfulness improvement over baseline LLMs and 93.23\% accuracy in RF pattern recognition tasks using CLIP-based vision models.
Lian et al. ~\cite{lian_configuration_2023} develop Ciri, a generic LLM-based configuration validation framework that leverages prompt engineering and few-shot learning to detect mis-configurations while addressing LLM hallucination and nondeterminism.

\subsubsection{Model tuning for hallucination mitigating}
At the model level, we categorize existing methods into finetuning, attention intervention, and decoding constraints.

\begin{figure}
    \centering
    \includegraphics[width=\linewidth]{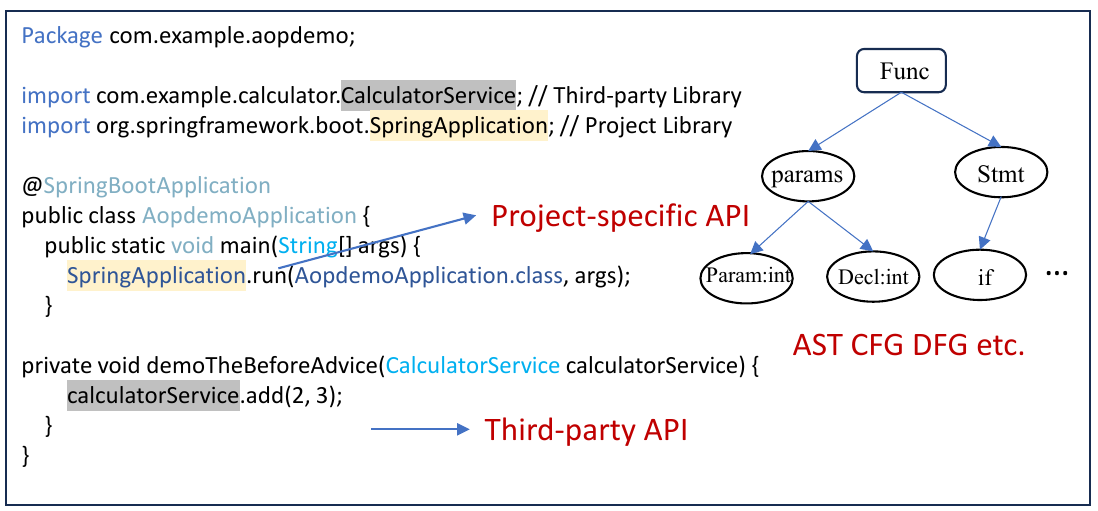}
    \caption{Challenges in Programming Languages: Domain-Specific APIs and Program Representation. }
    \label{fig:constrain}
\end{figure}

\textbf{Decoding Constraints. } Current studies ignore the structural dependencies in practical projects and do not validate whether the generated APIs are available or not. To address these limitations, Chen et al.~\cite{chen2025towards} propose MARIN, a framework for mitigating API hallucination in code generated by LLMs with hierarchical dependency awareness. 
Dong et al.~\cite{dong2025rethinking} formally define structural repetition and propose an efficient decoding approach named RPG (Repetition Penalization based on Grammar) to alleviate the repetition issues in LLM-based code generation.
Chen et al. ~\cite{chen2025towards} propose MARIN, a hierarchical dependency-aware framework that mitigates API hallucination in LLM-generated code through dependency mining and constrained decoding, significantly reducing hallucination metrics across multiple LLMs and real-world projects.

Sanchez et al. ~\cite{sanchez2024stay} demonstrate that Classifier-Free Guidance (CFG), originally from text-to-image generation, effectively enhances inference-time performance and prompt adherence across various language models and tasks, achieving state-of-the-art results.

\textbf{Retraining Approach. } This approach can be done through finetuning, reinforcement learning, etc.  Wang et al.~\cite{wang_insights_2025} apply reinforcement learning (RL), specifically direct preference optimization (DPO), to align Verilog code generation with functional correctness. 
Patil et al.~~\cite{patil_gorilla_2023} propose Gorilla, a fine-tuned LLaMA model trained with Retriever-Aware Training (RAT). Gorilla outperforms GPT-4 in API call generation, adapts well to test-time document changes, and significantly reduces hallucinations often seen in standard LLM prompting.
Nashaat et al. ~\cite{nashaat_towards_2024} propose CodeMentor, a few-shot learning framework combining weak supervision and reinforcement learning with human feedback to fine-tune LLMs for code review tasks, achieving significant improvements over state-of-the-art methods in code quality estimation, review generation, and bug report summarization tasks.
Dutta et al.~\cite{dutta_applying_2024} employ a specialized prompting strategy and train a reward model towards better alignment from smaller LLMs. 
Hou et al. ~\cite{hou_geocode-gpt_2025} introduce GeoCode-GPT-7B, the first LLM fine-tuned for geospatial code generation using novel domain-specific corpora and a comprehensive evaluation framework, demonstrating significant improvements in accuracy, readability, and executability across multiple geospatial coding tasks.

As shown in Fig.\ref{fig:constrain}, compared to natural language, decoding constraints in code involve unique challenges ~\cite{chen2025towards} due to its structural and contextual dependencies:

\begin{itemize}
    \item \textbf{Project-specific APIs.} Many hallucinations arise when the model fabricates or misuses project-defined APIs or variables that are only valid within a specific repository or context. Decoding such errors requires analyzing the project's repository and environment.

    \item \textbf{Third-party libraries.} Code generation often involves external dependencies. Models may hallucinate non-existent functions or misuse rarely documented APIs. Effective decoding must resolve imports and verify usage against real API documentation or available packages.

    \item \textbf{Version mismatch.} An API that exists in one version of a library may be deprecated or modified in another. Decoding may encounter errors unless the correct library version is verified during the process.
\end{itemize}

\subsubsection{Prompt/Reasoning for hallucination mitigating}
With the advancement of large language models, reasoning capabilities have progressively exhibited remarkable performance.
Techniques such as Chain-of-Thought and Tree-of-Thought are introduced to enhance critical thinking and reasoning. Although many models do not explicitly claim that these methods are designed to address hallucinations, they can mitigate such issues in practice effectively.
Hou et al.~\cite{hou_chain--programming_2024} propose the Chain of Programming (CoP) framework, which decomposes the code generation process into five steps: requirement analysis, algorithm design, code implementation, code debugging, and code annotation. 

\begin{figure}
    \centering
    \includegraphics[width=\linewidth]{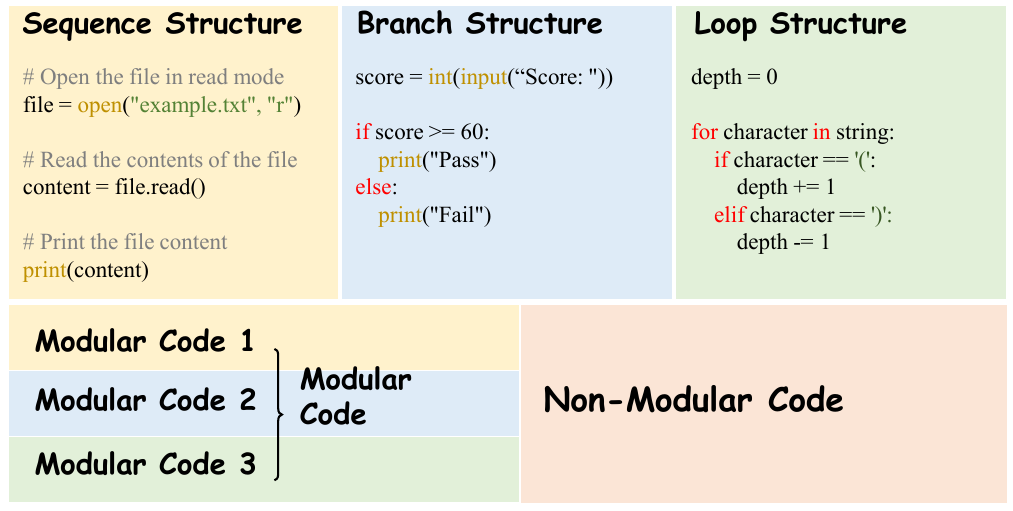}
    \caption{Fundamental Program Structures. }
    \label{fig:loop}
\end{figure}

Reasoning over code differs fundamentally from reasoning over natural language due to its strict syntax, structural hierarchy, and executable semantics, as shown in Fig.\ref{fig:loop}. For example, Code follows strict execution semantics. Structures like \texttt{if-else} and \texttt{for} loops enforce rigid logical flows that must be precisely preserved during reasoning. Any deviation leads to incorrect functionality, unlike the more flexible nature of natural language inference. Moreover, modularity facilitates localized reasoning by isolating functionality into well-defined components, while effective reasoning enables the correct interpretation, composition, and reuse of these modules. LLMs should perform both intra-module and inter-module reasoning to generate coherent and maintainable code. However, Kang et al.~\cite{kang2024revisiting} find that the modularity of a code example may not be the crucial factor for performance.

\subsubsection{Post-editing for hallucinations mitigating}
In addition to intervening in the model generation process, we also modify the content after the model has generated it, such as through self-refinement, iterative revision, and external validation. 

\begin{figure}
    \centering
    \includegraphics[width=\linewidth]{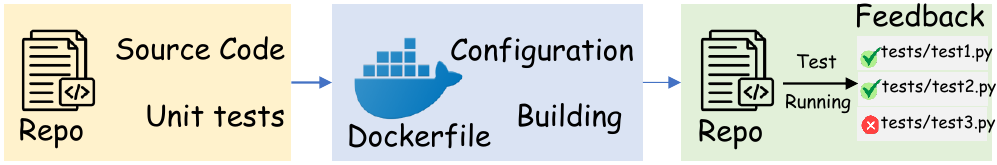}
    \caption{Code Configuration, Test, and Feedback. }
    \label{fig:code_test}
\end{figure}

Kouemo et al. ~\cite{ngassom_chain_2024} propose a self-refinement method using targeted verification questions on AST nodes to identify and repair bugs in LLM-generated code without test cases or human intervention, achieving significant error reduction and improved executability on the CoderEval benchmark.
Wang et al. ~\cite{wang_rat_2024} explore how iterative revising a chain of thoughts with the help of information retrieval significantly improves LLMs’ reasoning and generation ability in long-horizon generation tasks.
Xu et al. ~\cite{xu_understanding_2025} propose enhancing Web UI test repair by combining traditional matching techniques with ChatGPT’s language understanding and code generation, introducing an explanation validator to reduce hallucinations and improve repair accuracy on a benchmark dataset.
Maalek ~\cite{maalek_integrating_2024} proposes incorporating generative artificial intelligence LLMs into the Master of Science curriculum on digitization in construction, aiming to facilitate effective interaction with LLMs for code generation while also mitigating concerns such as bias and hallucination.
Ngassom et al. ~\cite{ngassom_chain_2024} propose a self-refinement method using targeted verification questions to automatically detect and repair bugs in LLM-generated code, improving reliability without requiring test cases or human intervention.
Cai et al. ~\cite{cai_automated_2025} propose Refine4LLM, a program refinement–based framework that guides LLMs to generate and verify correct code from formal specifications, addressing hallucination and opacity issues in end-to-end code generation.

\subsubsection{Agent-based approach for hallucination mitigation}
Mitigation of hallucinations can also be effectively accomplished by leveraging collaboration among multiple agents.

For example, Chen et al. ~\cite{chen_gamegpt_2023} propose GameGPT, a multi-agent collaborative framework for automating game development that addresses hallucination and redundancy in LLM-generated code through dual collaboration, layered lexicons, and decoupling strategies to enhance planning and code precision.
Mei~\cite{mei_gamevlm_2024-1} proposes a multi-agent framework, i.e., GameVLM, to enhance the decision-making process in robotic task planning.

In agent-based systems, reasoning over code differs significantly from reasoning over natural language due to the dynamic and executable nature of programming languages, as shown in Fig.\ref{fig:code_test}. When an agent is tasked with solving a coding problem or completing a programming task, it must not only understand the syntactic and semantic structure of code but also interact with the execution environment~\cite{hu2025llm}.
\begin{itemize}
    \item \textbf{Multi-phase Reasoning.} Code reasoning involves multiple stages, including code generation, compilation or interpretation, execution, and verification~\cite{chang2023self}. Unlike static language tasks, agents must consider dependencies, environment variables, and toolchains before code can be executed.

    \item \textbf{Environmental Constraints.} Agents must reason about the runtime environment, such as file systems, API availability, package versions, or even hardware settings, which may directly affect execution success. This contrasts with natural language tasks that typically require only contextual semantics~\cite{zhang2024codeagent}.

    \item \textbf{Executable Feedback.} During or after execution, the agent can receive concrete signals such as error messages, exceptions, test failures, or performance metrics. These feedback signals can guide iterative refinement, enabling the agent to revise its plan, fix hallucinated components, e.g., non-existent APIs, or optimize performance.

    \item \textbf{Iterative Planning and Reflection.} Agents often follow a loop of \texttt{Think $\rightarrow$ Code $\rightarrow$ Execute $\rightarrow$ Reflect}~\cite{jiang2024self}. Each iteration refines the reasoning process, leveraging execution results to update the internal model or retrieve more relevant information.

\end{itemize}

\begin{center}
    \begin{tcolorbox}
            \textbf{Summary of RQ3:} 
            \begin{enumerate}
            \item Most existing efforts focus on mitigating hallucinations through RAG, fine-tuning, and post-editing techniques, which are similar to NLP tasks.
            \item While generally effective, most of these approaches often overlook the unique structural and semantic properties of code.
            \item Solutions specifically tailored to code hallucination, such as those targeting variable misuse, API hallucinations, or structural inconsistencies, are still relatively scarce and remain an open research direction.  
            \end{enumerate}
    \end{tcolorbox}
\end{center}

%% file: sections/RQ4.tex
\section{RQ4: Tasks, Datasets, Models, and metrics} 

This section summarizes the benchmark settings used to evaluate hallucination in code intelligence, including the selected tasks, datasets, models, and evaluation metrics. We aim to understand how hallucinations manifest across different task types and model architectures, and how well current evaluation methods capture such phenomena.

\subsection{Tasks}

Tab.~\ref{tab:code-intelligence-tasks} provides a comprehensive overview of representative tasks in the field of code intelligence. These tasks cover a wide range of objectives, from basic capabilities like code completion and code generation to higher-level functionalities such as program synthesis, which generates complete programs from specifications, and test case generation, which focuses on validating program correctness~\cite{lu2021codexglue, allamanis2018survey}. Many of these tasks support software development and maintenance by automating reasoning over source code, natural language specifications, or both. For example, code summarization and code refinement enhance readability and quality, while bug detection and code repair ensure correctness and reliability. Meanwhile, tasks like clone detection and code retrieval facilitate software reuse and search in large repositories.

\begin{table}[ht]
\centering
\caption{Overview of Code Intelligence Tasks. }
\label{tab:code-intelligence-tasks}
\begin{tabularx}{\linewidth}{p{3cm}X}
\toprule
\textbf{Task} & \textbf{Description} \\
\midrule
Code Completion & Predict the next token, statement, or code block given a code prefix.  \\
Code Generation & Generate executable code from NL, specifications, or comments.  \\
Code Summarization & Generate concise NL summaries or comments for given code snippets or functions. \\
Code Translation & Translate code from one PL to another, such as Java to Python. \\
Code Retrieval & Retrieve relevant code snippets or APIs from a large corpus given a query, which could be code or NL. \\
Bug Detection & Automatically identify semantic or syntactic issues in code, such as vulnerabilities. \\
Code Repair & Automatically fix bugs or vulnerabilities in code based on context or known issues. \\
Clone Detection & Detect semantically or syntactically similar code fragments across codebases. \\
Type Inference & Predict missing or implicit type annotations, especially in dynamically typed languages. \\
Code Refinement & Improve code quality, structure, or efficiency without altering functionality, e.g., refactoring and optimization. \\
Test Case Generation &  Generate unit or integration tests based on the code implementation automatically. \\
Program Synthesis & Generate entire programs that meet a given specification, often under strict formal or functional constraints. \\
API Recommendations & Suggest relevant APIs or function calls based on NL intent, context, or code snippets.\\
\bottomrule
\end{tabularx}
\end{table}

\subsection{Datasets}

To systematically evaluate hallucination in code generation, we review a set of recent benchmarks that are specifically designed to expose factual inconsistencies, invalid API usages, and functional errors. These datasets cover diverse programming tasks, languages, and evaluation protocols, enabling both qualitative and quantitative assessment of hallucination phenomena in LLMs. Tab.~\ref{tab:hallucination-benchmarks} summarizes the key characteristics of these representative benchmarks, including their focus areas, e.g., hallucination detection, correctness verification, and security. These benchmarks serve as a foundation for measuring the factual grounding and reliability of code generated by LLMs across different scenarios.

Despite their valuable contributions, current benchmarks do not fully capture the multifaceted nature of hallucination in code generation. First, these benchmarks focus primarily on a limited set of PLs, e.g., Python or Java, leaving low-resource or domain-specific languages underrepresented. Second, while some datasets emphasize API-level hallucinations, others concentrate on functional correctness, with few offering unified annotations for both aspects. Moreover, hallucinations stemming from training data contamination, prompt sensitivity, or cross-domain generalization remain underexplored in code intelligence.

\begin{table}[ht]
\centering
\caption{Representative Benchmarks for Hallucination and Correctness Evaluation in Code LLMs. }
\label{tab:hallucination-benchmarks}
\begin{tabularx}{\linewidth}{llX}
\toprule
\textbf{Benchmark} & \textbf{Language} & \textbf{Description} \\
\midrule
CodeMirage~~\cite{agarwal2024codemirage} & Python & The dataset comprises 1,137 programming problems, each accompanied by LLM-generated hallucinated Python code snippets, reference (ground-truth) solutions, and associated test cases for systematic evaluation.
\\
CODEGUARD+~~\cite{fu2024constrained} & C++ and Python & Evaluates the ability of code LLMs to generate secure and correct code, with two new proposed metrics. \\
HALLUCODE~~\cite{liu2024exploring} & Python & Measures the recognition of hallucinations in code generation by LLMs. \\
CodeHaluEval~~\cite{tian2025codehalu} & Python & Contains 8,883 samples from 699 tasks, enabling systematic and quantitative evaluation of code hallucinations. \\
CoderEval~~\cite{yu2024codereval} & Python and Java & Includes 230 Python and 230 Java code generation tasks from real-world projects with an automatic execution framework to assess functional correctness. \\
\bottomrule
\end{tabularx}
\end{table}

\subsection{Code LLMs}

Tab.~\ref{tab:code-llms} summarizes representative LLMs specialized for code-related tasks. These models vary significantly in size, design philosophy, and training objectives. For example, Qwen2.5-Coder-32B-Instruct from Alibaba and DeepSeek-Coder-33B-Instruct from DeepSeek are both instruction-tuned models that support multilingual code generation and exhibit strong performance in both code completion and synthesis. CodeLlama-70B-Instruct and WizardCoder-34B are optimized for high accuracy on benchmarks such as HumanEval and MBPP, with the latter focusing specifically on Python. In contrast, StarCoder2-15B and CodeGeeX2-6B emphasize open-source availability and multilingual capabilities, supporting over 80 and 20 languages, respectively. Lightweight models like Replit Code V1.5-3B cater to efficiency and resource-constrained environments. 

\begin{table}[ht]
\centering
\caption{Representative of code LLMs.}
\label{tab:code-llms}
\begin{tabularx}{\linewidth}{lll>{\raggedright\arraybackslash}X}
\toprule
\textbf{Model} & \textbf{Parameters} & \textbf{Organization} & \textbf{Highlights} \\
\midrule
Qwen3-Coder-480B-A35B-Instruct~\cite{yang2025qwen3} & 480B/30B & Alibaba & Significant performance, long-context capabilities, agentic coding \\
Qwen2.5-Coder-32B-Instruct~\cite{hui2024qwen2} & 32B & Alibaba & Instruction-tuned, multilingual, supports various programming languages. \\
CodeLlama-70B-Instruct~\cite{roziere2023code}     & 70B & Meta    & Code-specialized LLaMA2 model, strong on HumanEval and MBPP. \\
StarCoder2-15B~\cite{lozhkov2024starcoder}             & 15B & BigCode & Trained on permissively licensed data, supports 80+ languages. \\
DeepSeek-Coder-33B-Instruct~\cite{guo2024deepseek} & 33B & DeepSeek & Multi-stage training; excels in code completion and generation. \\
WizardCoder-34B~\cite{luo2024wizardcoder}            & 34B & NL2Code & Python-focused, high accuracy on HumanEval. \\
CodeGen2.5-16B~\cite{nijkamp2023codegen2}             & 16B & Salesforce & Supports multilingual code generation, instruction-following. \\
replit-code-V1\_5-3B~\cite{replit2023}        & 3B  & Replit  & Lightweight, efficient, optimized for completions. \\
CodeGeeX2-6B~\cite{codegeex2_2023}               & 6B  & Tsinghua & Open-source, supports code translation among 20+ languages. \\
\bottomrule
\end{tabularx}
\end{table}

\subsection{Metrics}

To assess the quality of generated content, several commonly used automatic evaluation metrics, e.g., Exact Match (EM), CodeBLEU~\cite{ren2020codebleu}, and ROUGE~\cite{wei2019code}, are often adopted. While these metrics can reflect discrepancies between generated content and reference implementations, they are not designed to detect hallucination. Specifically, they primarily measure syntactic or lexical similarity, which may penalize semantically correct yet diverse solutions, and fail to capture factual inconsistencies such as the invocation of non-existent APIs.

\subsubsection{API Metrics}

In API-specific scenarios, two key elements in APIs are focused on, including the names and parameter patterns, and two metrics for evaluating the degree of API hallucination~\cite{chen2025towards}.

\textbf{Micro Hallucination Number (MiHN)} counts the average number of hallucinatory elements within the generated APIs:

\begin{equation}
\text{MiHN} = \frac{\sum_{i=1}^{n} \text{Count}(hallucinatory \ elements \ in \ a_i)} {n},
\end{equation}
where $n$ is the total number of generated APIs, and $a_i$ is the $i$-th generated API.

\textbf{Macro Hallucination Rate (MaHR)} calculates the proportion of generated APIs that contain any hallucinatory elements:

\begin{equation}
\text{MaHR} = \frac{\text{Count}(APIs \ with \ hallucinations)} {n},
\end{equation}
where $n$ is the total number of generated APIs.

\subsubsection{Functional Metric}

The \textbf{pass@k} metric measures the probability that at least one of the top-$k$ generated solutions for a problem passes all unit tests. 
\begin{equation}
\text{pass@}k := \mathbb{E} \left[ 1 - \frac{\binom{n - c}{k}}{\binom{n}{k}} \right]
\label{eq:passk}
\end{equation}

This formulation avoids the high variance of direct evaluation. A numerically stable implementation is recommended because of the large combinatorial values.

\subsubsection{Token Repetition at Phrase Level (TR-N)}
The \textbf{TR-N} metric captures structural repetition at the n-gram level~\cite{dong2025rethinking}. TR-N is defined as:
\begin{equation}
\small
\text{TR-N} = 1.0 - \frac{|\{ G(x)' \mid \exists p \in [1, |G(x)| - n + 1], G(x)' = G(x)_{p:p+n-1} \}|}{|G(x)| - n + 1}
\end{equation}
Here, $G(x)$ denotes the generated token sequence, and $n$ is the $n$-gram size. This metric reflects the proportion of duplicate n-grams in a sequence; lower values indicate more repetition~\cite{dong2025rethinking}.

\subsubsection{Token Repetition at Statement Level (TR-S)}

The \textbf{TR-S} metric evaluates repetition at the statement level and is defined as:
\begin{equation}
\text{TR-S} = 1.0 - \frac{\text{unique statements in } G(x)}{\text{statements in } G(x)}
\end{equation}
This measures the ratio of unique statements to the total number of statements in the generated output. Like TR-N, a lower TR-S indicates higher structural repetition.

\subsubsection{Variable Metrics}
Given two token sets \( S_A \) and \( S_B \) obtained from splitting variable names, the Jaccard similarity is defined as

\begin{equation}
\mathrm{Jaccard}(S_A, S_B) = \frac{|S_A \cap S_B|}{|S_A \cup S_B|}
\end{equation}

Given two token sequences \( A \) and \( B \), the Edit Similarity is defined based on the Levenshtein edit distance \( D_{lev}(A, B) \) as
\begin{equation}
\mathrm{EditSim}(A, B) = 1 - \frac{D_{lev}(A, B)}{\max(|A|, |B|)}
\end{equation}
where \( D_{lev}(A, B) \) denotes the minimum number of token insertions, deletions, or substitutions required to transform sequence \( A \) into \( B \).

\subsubsection{Compatibility Metrics}

\begin{equation}
\text{BuildSuccessRate} = \frac{|\{ c \in \mathcal{C} \mid \text{build}(c) = \text{true} \}|}{|\mathcal{C}|}
\end{equation}
\noindent
where, $\mathcal{C}$ denotes the set of generated code samples, and $\text{build}(c)$ returns \texttt{true} if the code sample $c$ can be successfully compiled or executed without errors. This metric reflects the model's ability to produce code that is syntactically correct, dependency-complete, and compatible with the target runtime environment.

\subsubsection{Classification-based Metrics}

Hallucination Rate (HR)~\cite{liu2024exploring}. HR is calculated after classification.

\begin{equation}
\mathrm{HR} = \frac{1}{N} \sum_{i=1}^{N} S(i, K),
\end{equation}
where $S(i, K)$ is an indicator function. If the $i^{th}$ sample satisfies the hallucination condition, then $S(i, K) = 1$; otherwise, $S(i, K) = 0$. Ideally, a lower HR indicates a greater robustness and reliability.

\subsubsection{LLM-as-a-judge Metrics}

The idea is to use evidence $E$ obtained from external sources to effectively verify whether a claim $c$ is hallucinated~\cite{wang2024halu}.
To operationalize this process, the HALU-J is proposed, a four-stage framework that categorizes, reorders, analyzes, and aggregates evidence to assess its relevance and quality for generating critiques or labels.

We formalize the \textit{LLM-as-a-judge} paradigm as the use of a large language model (LLM) to evaluate or compare candidate outputs given a task input. This setting can be represented as either a pointwise scoring function or a pairwise preference function.

\textbf{Pointwise Scoring.} Given a task input \( x \) and a set of candidate outputs \( \{y_1, y_2, \dots, y_k\} \), the LLM defines a scoring function:

\begin{equation}
   \text{Score}_{\text{LLM}}(x, y_i) \in \mathbb{R} 
\end{equation}

\textbf{Pairwise Preference.} Alternatively, the LLM can perform comparative evaluation by selecting a preferred output between two candidates:

\begin{equation}
\text{Judge}_{\text{LLM}}(x, y_i, y_j) =
\begin{cases}
y_i \succ y_j & \text{if } y_i \text{ is preferred over } y_j \\
y_j \succ y_i & \text{otherwise}
\end{cases}
\end{equation}

\begin{center}
    \begin{tcolorbox}
            \textbf{Summary of RQ4:} 
            \begin{enumerate}
            \item Most code hallucination datasets focus on a few languages, e.g., Python and Java, limiting their broader applicability.
            \item Metrics applicable to the hallucination scenario include syntax validity, functional correctness, exact match, and hallucination-specific scores, e.g., MiHN, Pass@K, and HR. 
            \end{enumerate}
    \end{tcolorbox}
\end{center}

%% file: sections/chanllenges.tex
\section{Challenges and opportunities}
\begin{figure}
    \centering
    \includegraphics[width=0.44\linewidth]{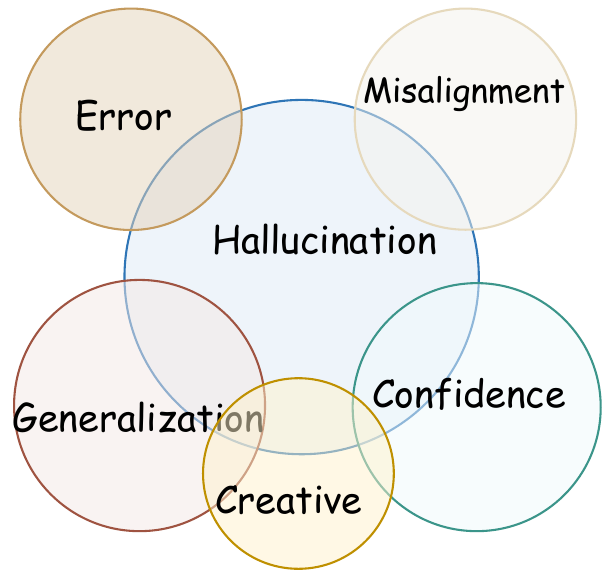}
    \caption{Conceptual Landscape of Hallucination in AI Systems.}
    \label{fig:relation}
\end{figure}

As LLMs continue to advance in code-related capabilities, new challenges emerge alongside new opportunities. In particular, the phenomenon of hallucination poses significant threats to the reliability, safety, and usability of generated code. Unlike NL, which is primarily intended for human interpretation, PLs must be correctly understood by both humans and machines~\cite{buse2009learning,allamanis2018survey}. This dual requirement makes the code intelligence tasks particularly sensitive to hallucinations introduced by LLMs. In addition, many concepts in AI systems are difficult to distinguish, as shown in Fig.\ref{fig:relation}, and we discuss a selected subset of these concepts.

\subsection{Challenges}

\subsubsection{Differentiating hallucination from mistake}

Hallucinations are a type of output error produced by deep neural networks. While often used interchangeably, \textit{hallucination} and \textit{mistake} refer to different phenomena in the behavior of LLMs~\cite{liu2023your}. A \textbf{mistake} typically denotes an error in syntax, logic, or factual recall that can be attributed to insufficient learning, ambiguous input, or random variation. In contrast, a \textbf{hallucination} refers to output that is fluent and syntactically plausible, yet factually incorrect or unsupported by the input or any underlying knowledge. For instance, if a model generates pritn("Hello") instead of print("Hello"), this is a mistake due to a syntactic slip; but if the model confidently produces a function calculateSquareRoot() while no such API exists in the referenced library, this is a hallucination. Hallucinations are especially problematic because they are often produced with high confidence, making them harder to detect and more misleading than simple mistakes. As such, hallucinations reflect deeper issues of model faithfulness and factual grounding, rather than surface-level performance errors.

\begin{table}[ht]
\centering
\caption{Comparison between Mistake and Hallucination in Language Model Outputs}
\label{tab:mistake_hallucination}
\renewcommand{\arraystretch}{1.3}
\begin{tabular}{@{}p{3cm} p{5cm} p{5cm}@{}}
\toprule
\textbf{Aspect} & \textbf{Mistake} & \textbf{Hallucination} \\ \midrule
Definition & A general error in syntax, logic, or factual recall due to ambiguity or insufficient learning & A fluent but factually incorrect output not supported by the input or underlying knowledge \\
Source & Often linked to weak learning signals or ambiguous inputs & Generated without grounding in the input \\
Form & Spelling error, syntax error, wrong logic, incorrect answer & Invented facts, made-up functions, unsupported claims \\
Model Confidence & May be low or uncertain & Often produced with high confidence \\
Detectability & Relatively easy to spot and fix & Harder to detect; can appear plausible and misleading \\
Impact & Limited, often superficial & Severe, especially when used in high-stakes tasks \\ \bottomrule
\end{tabular}
\end{table}

\subsubsection{Understanding the Relationship Between Model Confidence and Hallucination}

Although language models often assign high probabilities to generated sequences, such high confidence does not necessarily imply factual correctness. Let \( P(y \mid x) \) denote the model’s conditional probability of generating output \( y \) given input \( x \); a high value of \( P(y \mid x) \) or a low perplexity score, defined as

\begin{equation}
    \mathrm{PPL}(y \mid x) = \exp\left(-\frac{1}{n} \sum_{i=1}^n \log P(y_i \mid y_{<i}, x)\right),
\end{equation}

The PPL only indicates the model’s internal certainty based on training distributions, not the factual validity of \( y \). Consequently, the model may produce hallucinations—outputs that are syntactically fluent but semantically or factually incorrect, even when \( P(y \mid x) \) is high. This phenomenon, often referred to as \emph{confident hallucination}, suggests that confidence scores and perplexity are insufficient for evaluating factual faithfulness, and calls for additional external verification methods.

\subsubsection{Balancing Hallucinations with Generalization}

It is widely acknowledged that hallucinations are a consequence of a balance between creativity and factual accuracy, which can be mitigated, but not eliminated~\cite{li2024banishing}. Li et al.~\cite{li2024banishing} find that models with low generalization error can still hallucinate significantly. Wong et al.~~\cite{wong2025generalization} present tensor completion and artifacts as a unified framework for explaining specific types of hallucinations and generalizations in language models.

\begin{table}[ht]
\centering
\caption{Comparison between Generalization and Hallucination in Language Models}
\label{tab:gen_hallu}\textbf{}
\renewcommand{\arraystretch}{1.3}
\begin{tabular}{@{}p{3cm} p{5cm} p{5cm}@{}}
\toprule
\textbf{Aspect} & \textbf{Generalization} & \textbf{Hallucination} \\ \midrule
Definition & Producing correct outputs on inputs not seen during training & Producing plausible but factually incorrect or unsupported outputs \\
Goal/Effect & Desirable model behavior enabling adaptability and robustness & Undesirable behavior that reduces factuality and trustworthiness \\
Grounding in Input & Output is supported or logically entailed by the input & Output may contradict or go beyond the given input \\
Relation to Training Data & Extends learned patterns appropriately & May fabricate information not present in the training data or infer content without a reliable basis \\
Evaluation Perspective & Success is measured by accuracy and generalization error & Evaluated by factual consistency and faithfulness to the input \\
Trade-off View & Encouraged for improved model performance & Needs to be minimized for reliability \\
Co-occurrence & Can co-exist with low hallucination if well controlled & May occur even with low generalization error \\ \bottomrule
\end{tabular}
\end{table}

\subsubsection{Absence of design pattern awareness}

Current code LLMs often lack explicit understanding or application of software design patterns. While some common patterns may be learned implicitly from training data, models rarely generate code that adheres to structured architectural principles. This deficiency not only limits the modularity, maintainability, and scalability of generated code but also increases the likelihood of structural hallucinations, where the model produces syntactically correct code that violates high-level design intent or misuses architectural patterns. Such hallucinations are harder to detect than low-level syntax errors, as they compromise software quality at a more abstract level. Addressing this challenge should involve pattern-aware prompts, architectural constraints, or retrieval-augmented generation to guide the model toward producing well-structured, semantically aligned code.

\subsubsection{Factuality and hallucination in synthetic data}

While synthetic data offers a promising solution to address the challenges of data scarcity and privacy concerns, a critical concern lies in ensuring its factuality and fidelity~\cite{liu2024best}. Models trained on synthetic datasets that contain hallucinated, false, or biased information may fail to generalize effectively to real-world scenarios, leading to degraded performance and reliability. This risk underscores the importance of rigorous data validation, bias mitigation, and the development of robust generative frameworks.

\subsubsection{Preventing hallucinations that expose private or sensitive data}

One critical yet underexplored risk of hallucinations in large language models is the potential exposure of private or sensitive data. 
Industrial code exhibits significant distributional shifts from open-source data, while privacy constraints preclude the use of commercial solutions, posing unique challenges for code completion in real-world settings~\cite{wang2025rag}.
When models are pre-trained on publicly available code repositories, documentation, or user-generated content, they may memorize and reproduce unsuitable code snippets, API, or internal project identifiers.

\subsection{Opportunities}

\subsubsection{Unified evaluation benchmarks}

Current benchmarks covered are relatively limited, with most benchmarks concentrating on a narrow set, such as Python and Java, lacking other widely used or domain-specific languages, e.g., Rust, or DSLs. Another significant opportunity is to construct dynamic updates to the dataset in response to emerging coding practices and user needs. Because static design not only limits the relevance and realism of evaluations over time but also increases the risk of data leakage, as LLMs may be pre-exposed to fixed benchmark datasets during training~\cite{li2024evocodebench}.
Building unified, up-to-date benchmarks that include private repositories and standardized evaluation metrics would enable more consistent comparison and better tracking of progress across domains.

\subsubsection{Explainable hallucination detection}

Most existing methods treat hallucination as a classification problem. Developing interpretable models that can explain why and how hallucinations occur would improve trust and usability.
One promising direction is to analyze the interaction between input prompts and generated outputs by identifying the combinations of input tokens that most strongly influence the predictions~\cite{chang2024xprompt}. Such approaches enable counterfactual explanations, for instance, by systematically modifying or removing specific parts of the prompt to observe whether and how the hallucinated content changes. This process offers valuable insights into the sources and triggers of hallucinations, thereby informing more effective mitigation strategies.



\subsubsection{Attribution through model editing}
Model editing, modifying specific behaviors of an LLM without retraining, can serve as a tool for attribution by identifying and isolating the
representations for hallucinations~\cite{hartvigsen2023aging}. By observing how edits propagate across tasks and outputs, researchers can trace the model’s reliance on erroneous or outdated knowledge. This enables more targeted hallucination mitigation and offers insights into the internal structure of code-specific knowledge within LLMs.

\subsubsection{Unified multi-stage mitigation pipelines}

Existing strategies often target a single stage, e.g., pre- or post-generation. Designing unified frameworks that combine code retrieval, structure-aware decoding, and post-generation verification may offer more robust and interpretable mitigation. For example, knowledge-enhanced retrieval addresses information gaps upstream, while constraint-based decoding mitigates structural hallucinations midstream, and symbolic execution helps validate outputs downstream. Such modular composition improves robustness and provides a foundation for explainable hallucination mitigation.

\subsubsection{Continual learning and lifelong adaptation}

Codebases and APIs evolve, and outdated knowledge is a key source of hallucinations in code generation. Continual learning enables LLMs to adapt to evolving software ecosystems without full retraining incrementally. By incorporating updates from repositories, API changes, and usage patterns, models can remain aligned with current development practices. Such lifelong adaptation is crucial to reducing hallucinations caused by stale or deprecated information, especially in rapidly evolving software domains.

\subsubsection{Mitigating hallucinations in multimodal UI code generation}

Multimodal large language models (MLLMs) show great promise in UI code generation by translating inputs into executable code. To mitigate hallucinations in this field, grounding mechanisms can be introduced to ensure that all generated elements have a corresponding representation in the visual or textual input. Incorporating design constraints, such as layout grids, style tokens, and component libraries, during training or decoding further reduces speculative generation. In addition, real-time feedback from rendering environments or static UI validators can provide a feedback loop~~\cite{fan2025user}, allowing the model to self-correct hallucinated elements by verifying output plausibility. 

\subsubsection{Model merging for complementary strengths}

Merging multiple LLMs with complementary strengths offers a promising direction for reducing hallucinations in code generation. For instance, combining a general-purpose LLM with a smaller and code-specific LLM can help balance linguistic fluency with structural correctness~\cite{tam2024llm}. Model merging techniques, e.g., weight interpolation, low-rank adaptation fusion, or modular ensembling, can preserve the factual grounding of one model while leveraging the generation capabilities of another. This approach can enhance robustness and reduce hallucination by aligning outputs with multiple learned distributions.

%% file: sections/discussion.tex
\section{Threats to validity}
This paper aims to provide a comprehensive and up-to-date overview of hallucination phenomena in LLM-based code intelligence. We systematically review literature from the past three years, a period marked by rapid advancement in LLM capabilities and increasing interest in code generation tasks. Our goal is to capture recent trends, categorize mitigation techniques, and highlight open challenges and future opportunities. However, as with any systematic review, certain threats to validity can not be fully eliminated, particularly in the areas of literature selection and data analysis.

\subsection{Literature Selection Validity}
Literature selection concerns the completeness and representativeness of the surveyed studies. In this paper, we focus on publications related to code hallucinations, defined as factually incorrect, illogical, or non-executable outputs produced by LLMs during code generation or reasoning. Since this is an emerging research area, there is no standardized terminology or taxonomy, and relevant work may appear under related but diverse keywords such as ``code hallucination'', or ``code faithfulness''. To address this problem, we collect papers using combinations of relevant keywords across three major academic databases, e.g., ACM Digital Library, IEEE Explore, and Web of Science. In line with our goal of surveying recent developments, we limit the scope to publications from 2022 to 2025, which may result in the omission of some earlier foundational work. 
The paper selection is carried out collaboratively: the first two authors applied predefined inclusion and exclusion criteria, and the remaining authors independently verified the results to minimize potential bias. Nonetheless, the possibility of missing relevant studies remains a moderate threat.
In addition, to further highlight the unique characteristics of code intelligence, we also compile representative papers from this domain.

\subsection{Data Analysis Validity}

The threat to data analysis validity arises from potential errors in extracting, interpreting, or classifying information from the surveyed papers. Given the limited number of publications in this emerging domain, inconsistencies in terminology, evaluation protocols, and definitions of hallucination may introduce ambiguity. To mitigate this, we adopt a two-stage data extraction strategy. Two independent annotators review each selected paper and extract key metadata, e.g., hallucination type, task domain, and mitigation technique, which are then cross-validated by a third reviewer. Disagreements are resolved through group discussions until a consensus is reached. While we strive to ensure consistency and accuracy, the risk of mis-classification or subjective interpretation remains, especially in cases where papers lack explicit descriptions. Moreover, as evaluation metrics for hallucinations are still evolving, our comparisons may reflect limitations inherent in the original studies.

%% file: sections/conclusion.tex
\section{Conclusion}

In this survey, we provided a comprehensive examination of hallucination in code-oriented Large Language Models  (LLMs) from four key perspectives. 
We examine two sets of papers: (1) NLP surveys that summarize hallucination research in natural language generation, and (2) software engineering papers that directly investigate hallucinations in code. By comparing insights from these two perspectives, we manually define and characterize hallucination in code, identifying its primary causes, including data noise, exposure bias, and insufficient semantic grounding, and analyze its evolving treatment across the NLP and software engineering communities. We summarize a range of mitigation strategies, such as knowledge-enhanced decoding, constrained generation, and consistency training, many of which originate from NLP research. In addition, we highlight code-specific challenges that intensify hallucination, including syntax rigidity, strong typing systems, and external library dependencies, and discuss how program analysis, symbolic execution, and unit testing are being adopted to detect and alleviate hallucinated outputs. Lastly, we review current evaluation methodologies and stress the importance of developing unified, hallucination-aware benchmarks that go beyond general correctness.
We conclude by identifying several open challenges and future directions. These include balancing generalization with factual reliability, and addressing hallucination detection in private codebases. 